\newcommand{\ignore}[1]{}
\documentstyle[11pt,epsf]{article}
\newcommand\be{\begin{equation}}
\newcommand\ee{\end{equation}}
\newcommand\bea{\begin{eqnarray}}
\newcommand\eea{\end{eqnarray}}

\newcommand{\smfrac}[2]{{\textstyle{{#1}\over{#2}}}}

\newcommand{\E}[1]{e^{\textstyle {#1}}}

\begin{document}
\begin{titlepage}
\noindent BU-CCS-950601 \\ BROWN-HET-999 \hfill hep-lat 9509012
\\MIT-CTP-2461  \\

\begin{center}
{\Large\bf Chronological Inversion Method for the Dirac Matrix\\
\bigskip
in Hybrid Monte Carlo}

\vspace{1.0cm}
{\bf R.C.~Brower$^{(1)}$, T.~Ivanenko$^{(2)}$,  
A.R.~Levi$^{(1)}$ and K.N.~Orginos$^{(3)}$}
\end{center}

\vspace{0.5cm}
\begin{flushleft}
{}~~$^{(1)}$Department of Physics, Boston University, Boston, MA
02215, USA \\ {}~~$^{(2)}$Center for Theoretical Physics,
Mass. Inst. of Tech., Cambridge, MA 02139, USA \\
{}~~$^{(3)}$Department of Physics, Brown University, Providence, RI
02912, USA\\
\end{flushleft}
\vspace{1.0cm}

\abstract{ 
  In Hybrid Monte Carlo simulations for full QCD, the gauge fields
  evolve smoothly as a function of Molecular Dynamics time.  Here we
  investigate improved methods of estimating the trial or starting
  solutions for the Dirac matrix inversion as superpositions of a
  chronological sequence of solutions in the recent past.  By taking
  as the trial solution the vector which minimizes the residual
  in the linear space spanned by the past solutions, the number 
  of conjugate gradient iterations per unit MD time is decreased 
  by at least a factor of 2. Extensions of this basic approach to 
  precondition the conjugate gradient iterations are also discussed.  }
\vfill
\end{titlepage}

{\bf I. Introduction}

At present, typical methods for going beyond the quenched (or valence)
approximation for lattice QCD involve an increase of several orders of
magnitude in computational time --- severely limiting statistics on
large lattices. Consequently, finding a more efficient approach to the
inclusion of internal fermion loops poses a major challenge for the
next generation of lattice simulations~\cite{Teraflops}.

The most successful approach to date for generating full QCD
configurations is the so-called Hybrid Monte Carlo (HMC)
algorithm~\cite{HMC}, which uses Molecular Dynamic (MD) evolution in a
``fifth time'' coordinate t.  At each time step, the Dirac matrix must
be inverted to calculate the force due to fermion loops.  This
inversion is the most time-consuming part of the full QCD
algorithm. The Dirac operator is represented as a $12 V$ by $12 V$
sparse square matrix, where the number of space-time lattice sites (or
volume $V$) is on the order of $10^{5-6}$. With the lattice volumes
used in recent simulations, the Dirac matrix may account for 90\% or
more of the computer's time.  Moreover, due to the Dirac inverter, the
problem scales as $(1/a)^{6-7}$.  Therefore any acceleration of the
Dirac inversion implies a major improvement in the overall performance
of generating QCD configurations.

Our simple observation is that by virtue of the system's smooth
evolution in MD time, it should be possible to use information from
the recent past to perform the next inverse more
efficiently~\cite{sant}.

Our present approach to make use of the past data is
straightforward~\cite{blk}.  Since the conjugate gradient is an
iterative procedure, a trial starting configuration is needed.  Our
method is to ``extrapolate'' from a set of the most recent solutions
of the Dirac inversion in order to find a better trial solution for the next
inversion.  This idea is legitimized by the observation, that if the
inverse is solved exactly, detail balance is preserved, regardless of
the starting trial solution for the conjugate gradient. 
Therefore we do not propose
to modify the QCD algorithm, but only to provide a prescription for
better selecting the starting trial solution for the Dirac inverter.

This method, like the interesting suggestion of
L\"uscher~\cite{Leuscher}, relies on the observation that the new
generation of supercomputers often has very large memories and that
the efficient use of the entire resource is no longer simply the
optimization of the CPU but also of its memory and its communication
systems. In particular, improved algorithms based on the exploitation
of a large memory are to be expected, since most QCD algorithms have
been designed at a time when memory was at a premium.

Several possibilities can be taken into account.  The first obvious
step, employed frequently in HMC codes, is simply to use the previous
solution as the starting trial solution. Some groups have even used a
linear extrapolation of the last two solutions\cite{gltrs,Gupta}.
The next natural step is to use a high order polynomial
extrapolation~\cite{blk}. We analyzed the polynomial extrapolation up
to the sixth order and we have observed that it gives a robust speed up in
the CG (see section V).  Motivated by this success, we then considered
the possibility of fixing the trial vector as a linear combination of
the $N$ previous solutions, by minimizing the residual in the norm of
the matrix (see section IV). This approach, which we call the Minimal
Residual Extrapolation (MRE) method, further improved the performance.  We
also explored other possibilities, such as the Kalman
filter~\cite{Kalman}, however, in our simulations, none of these
methods offers better performance than the Minimal Residual Extrapolation
method.

Of course, as discussed in the conclusions, there may well be more
sophisticated ways to use past information, which will lead to a still
more efficient algorithm.  For example, it appears to us that a
sensible strategy is to exploit the past Krylov spaces to build
preconditions for the conjugate gradient iteration itself. As an
illustration, we present a modest step in this direction which builds
a preconditioner from the same vectors used for the MRE trial solution
(see Section VII.). We gain an additional small improvement in
performance, which encourages us to direct our future research toward
chronological preconditioners.

It is also noteworthy that this generic problem of a chronological
sequence of matrix inversions is not unique to HMC for QCD. For
example, in many fluid dynamics codes, one must solve Poisson's
equation for the pressure field as an inner loop for integrating the
Navier Stokes equation~\cite{NavierStokes}.  Consequently, there will
be other circumstances for developing and testing these algorithms.

\vskip 0.7cm 

{\bf  II. Hybrid Monte Carlo for QCD}

In this section, a brief review of the Hybrid Monte Carlo (HMC)
algorithm for full QCD is presented.  This algorithm has the advantage
that, apart from numerical round off, it provides an exact Markov
process for generating QCD configurations.  The starting point is the
Euclidean partition function for QCD,
\begin{equation}
        Z_{QCD}~=~\int D\psi D\bar\psi DU \E{- S_g(U) +\bar\psi M(U) \psi} \ ,
\label{zermelo}
\end{equation}
where the gauge variables are represented by unitary matrices
$U_\mu(x)=\E{i A_\mu(x)}$ and the Fermions by Grassmann
variables $\psi_x$ -- assigned to links and sites respectively.
$S_g(U)$ is the pure gauge action and $M(U)$ is the Dirac matrix,
\begin{equation}
M_{x,y} = \delta_{x,y} -  
      \kappa (1- \gamma_\mu) U_\mu(x) \delta_{x+\mu,y} 
      -  \kappa (1+ \gamma_\mu) U^\dagger_\mu(x) \delta_{x,y+\mu}
                      \; ,
\label{eq:wilson}
\end{equation}
where $\kappa$ is the hopping parameter.

We restrict our attention to 2 flavors of Wilson Fermions, so that
integrating out the Grassmann variables yields the square of the
Dirac determinant. Consequently, with the identity
\begin{equation}
       \biggr[  \det(M) \biggl]^{nf}~
        =~\biggr[  \det(M^\dagger M)~ \biggl]^{nf/2},
\label{marco}
\end{equation}
we may rewrite the determinant as integral over a positive definite Gaussian 
in terms of a set of bosonic ``pseudo-fermion'' fields $\varphi_x$.
The HMC algorithm also requires additional canonical (angular) momenta
coordinates, $E_\mu(x)$, conjugate to the gauge fields $A_\mu(x)$ on
each link $(x, x+\mu)$.  Combining these steps the QCD partition
function is now rewritten as
\begin{equation}
        Z_{QCD}~=~\int DU~DE~D\varphi ~\E{- H(U,E,\varphi)} \ ,
\end{equation}
with an entirely bosonic action given by 
\begin{equation}
         H(U,E,\varphi) = \smfrac{1}{2} Tr~E^2 + S_g(U) + 
           \varphi^\dagger [ M^\dagger M ]^{-1}\varphi \ .
\end{equation}
The standard HMC algorithm uses alternates Gaussian updates with
Hamiltonian evolution to achieve detailed balance and ergodicity: At
the beginning of each Hamiltonian trajectory the momenta, $E$, and
``pseudo-fermion'' fields, $b = {M^\dagger}^{-1} \varphi$, are chosen
as independent Gaussian random variables.  Next the gauge fields are
evolved for MD time $T$ using the Hamiltonian equations of motion for
fixed values of the``pseudo-fermion'' fields $\varphi$. It is easy to
prove that this results in Markov process for the gauge fields $U$
which leaves probability distribution,
\begin{equation}
{\cal P}(U) = ~\int D\psi D\bar\psi \E{- S_g(U) +\bar\psi M(U) \psi} \ ,
\label{eq:probability}
\end{equation}
invariant. 

In practice, the Hamiltonian evolution must be approximate by an
integration scheme with finite step size $\delta t$. Consequently, at
the end of each MD trajectory, a Metropolis accept-reject test is
introduced to remove integration errors which would otherwise corrupt
the action being simulated.  In addition, it is also essential that
the MD integration method exactly obey time reversal invariance to
avoid violating detailed balance.  Clearly small violations of
detailed balance are inevitable due to round off errors. Current
practice usually employs the leapfrog integration scheme in single
precision arithmetic.  Also, the standard odd-even
preconditioning~\cite{Gupta,DeGrand}, has been implemented by the
replacement,
\begin{equation}
M = 1 -\kappa K_{eo} - \kappa K_{oe} \; 
    \rightarrow  \; M_{ee} = 1 - \kappa^2  K_{eo} K_{oe} \ .
\end{equation}  
In this case, only pseudo-Fermions $\varphi$ on the even sites have to be
introduced and the matrix,
\begin{equation}
A = M^\dagger_{ee} M_{ee} \ ,
\end{equation}
is used in the algorithm.

There are many small variations on this basic method, but common to all HMC
algorithms is the need to accurately integrate the equations of
motion, calculating the force on $U$ due to the pseudo-Fermions at each
time step $t_n$. The force of the fermion loops on the
gauge field requires solving the Dirac equation,
\begin{equation}
      A(t_n)~ \chi(t_n) ~=~ \varphi \ ,
\label{eq:linear}
\end{equation}
over and over again for the propagator $\chi(t)$, where $A(t)\equiv
M(U)^\dagger M(U)$. Technically, this is achieved by starting with a
trial value $\chi_{trial}$ and iteratively solving 
Equation (\ref{eq:linear})
for $\chi(t)$. The operator A(t) changes smoothly as a function of the
MD time t, as new values of U are generated. During the MD steps for
which $\varphi$ is held fixed, the solution $\chi(t)$ must also 
change smoothly in time t.  
The solution of (\ref{eq:linear}), which is usually 
obtained using the conjugate
gradient (CG) method, is the most computationally expensive part of
Hybrid Monte Carlo algorithm.  Therefore improvements in this
part will result in nearly a proportional net gain in the performance of 
the full QCD simulations.

\vskip 0.7cm 

{\bf  III.  Simulations}

We have investigated the performance of our methods using the MIT-BU
collaboration code for the CM-5.  We have available two CM-5, one with
128 nodes and the other with 64 nodes. We used the configurations
generated for the MIT-BU lattice collaboration.  We tested various
values of the molecular dynamics step $\delta t$ for 5 or more
independent thermalized full QCD configurations, separated by
approximately 100 MD trajectories.

In our tests we performed extensive simulations on a $16^4$ lattice at
$\beta = 5.5$ and $\kappa = 0.160$.  We choose $\delta t$ in a typical
window for actual simulations ($0.002 \le \delta t \le 0.015$).  Some
simulations were performed with a lighter quark mass and we obtained
similar results~\cite{blk}, although the magnitude of the improvement
depended slightly on the lattice parameters.

We used a standard CG method, and we iterate  until the normalized
squared residual,
\begin{equation}
R = \frac{| M^\dagger_{ee} M_{ee} \chi - \varphi|^2}{|\chi|^2} \ ,
\label{eq:stopping}
\end{equation}
reaches a given value. We choose the stopping condition, in much the
same spirit as reference~\cite{Gupta}, so that the error in
computing the $\Delta S$, used in the Metropolis accept-reject step,
should on average be less than 1\%.  However it is not really known
how this error propagates to physical quantities.

\begin{figure}
$$
\epsfxsize=15.5cm
\epsfysize=13.0cm
\epsfbox{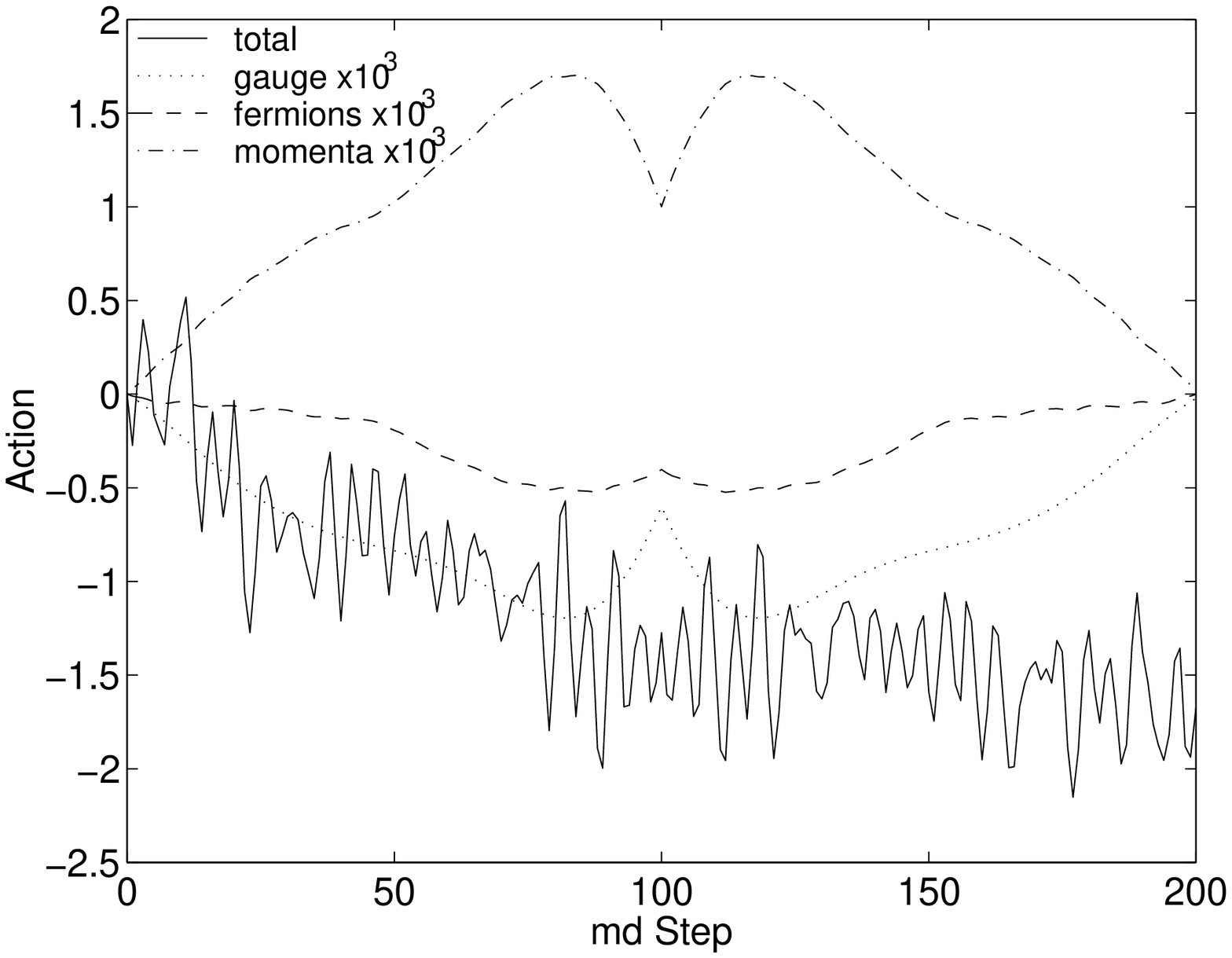} 
$$
\vspace{-1.0cm }
\caption{ Action variation in a forward-backward trajectory. 
    The initial action has been subtracted. 
    $R<10^{-10}$ was used as a stopping criterion and
    $\delta t = 0.010$.
\label{fig:reversible10} }
\end{figure}

\begin{figure}
$$
\epsfxsize=15.6cm
\epsfysize=19.5cm 
\epsfbox{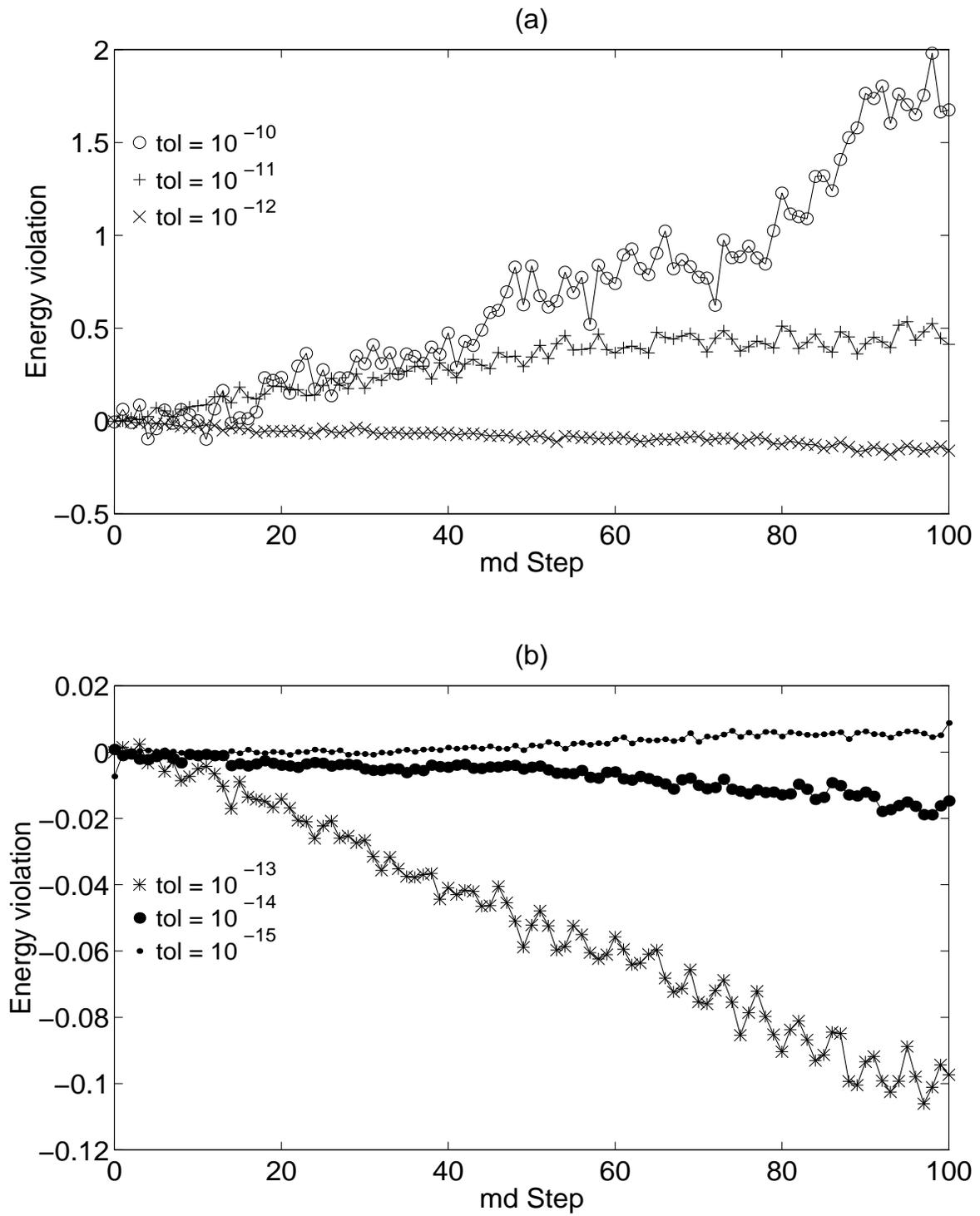} 
$$
\vspace{-1.0cm}
\caption{Reversibility violations for the energy differences in 
a  forward-backward trajectory with the stopping conditions 
$R = 10^{-10},10^{-11},10^{-12} $ (a) 
and $R = 10^{-13},10^{-14},10^{-15} $ (b).
\label{fig:reversible} }
\end{figure}

A critical issue is the requirement to converge accurately to the
final solution for $\chi$. Unless you start from a value of $\chi$
which is {\it independent } of past values (e.g. $\chi = \varphi$),
the initial guess introduces an element of non-reversible dynamics and
therefore a failure of detailed balance.  Even in one of the most
commonly used HMC methods, starting from the last value of $\chi$ (or
a constant extrapolation), failure to converge to the correct inverse
introduces a bias that breaks time reversal invariance.  In principle,
only if the CG has converged exactly to the fixed point of the
conjugate gradient iterations, there will be no violation of time
reversal invariance. In practice, we have measured this source of
error by explicitly reversing the MD dynamics for various values of
the stopping conditions (see Eq.~\ref{eq:stopping}).  In
Figures~\ref{fig:reversible10},\ref{fig:reversible} we show examples
for $R = 10^{-10}$, $R = 10^{-11}$, $R = 10^{-12}$, $R = 10^{-13}$, $R
= 10^{-14}$ and $R = 10^{-15}.$ In Figure~\ref{fig:reversible10} we
can see the action variation (total, fermion, gauge and momenta) in a
trajectory that goes forward in MD time 100 steps and then is reversed
for 100 steps.  We have subtracted the initial action, so if the
dynamics conserves energy, the total action should remain zero
and if the dynamics is reversible, the total action should be
symmetric around the MD time $t = 100$.

Our data demonstrates that simulations with $R \ge 10^{-12}$ are not
sufficient to ensure a reversible dynamics.  In
Figure~\ref{fig:reversible}, we plot the total action difference
between symmetric points in a forward-backward trajectory.  If the
dynamics is reversible, this difference should be zero.  In this
figure we can see that for $R>10^{-13}$ the dynamics is not
reversible. For $R=10^{-13}$ the action difference observed is of
order of $10^{-2}$ where the total action is of order $10^7$, which
means that we are at the boundaries of single precision accuracy.
However, for  $R=10^{-14}$ we are well bellow the single precision accuracy.
Smaller action variations can not be detected. Consequently, we chose
to run our simulation using $R=10^{-14}$ as the conjugate gradient
stopping criterion.

We measured the total number of CG steps needed to invert the Dirac
Matrix to a given precision.  The CPU time for a single
leap-frog trajectory is practically proportional to the number of CG
steps. However, the true performance of the method should be judged by
the total number of CG steps needed to evolve the system for a fixed
MD time, i.e. the total number of CG steps needed to compute a
trajectory of time length $T$.  We define the quantity,
\begin{equation}
    CT =  { N_{CG} \over \delta t } \; ,
\end{equation}
which is proportional to the ``computational time''.  Moreover, since
a smaller step size will improve the acceptance rate, the trade-off
may be even more favorable for small $\delta t$, and this effect
should also be included in an overall measure of efficiency.  However,
because we always worked with $\delta t$ in a region of very good
acceptance rates ($>99\% $), we have not included this effect in our
estimate of overall efficiency in our simulations.

\vskip 0.7cm 

{\bf  IV.  Chronological Inverter for QCD}

We began our investigation by considering a trial solution, $\chi(0)$, 
of the new Dirac matrix $A(0)$, as a linear superposition of old solutions,
\begin{equation}
       \chi_{trial} = c_1 \; \chi(t_1) + c_2 \; \chi(t_2) 
        + \cdots + c_N \; \chi(t_N). 
\label{timecoef}
\end{equation}
To simplify our notation, we will always suppose that the new inverse
is computed at $t = 0$, given the past values at $t_i$, with $\cdots <
t_3 < t_2 < t_1 < 0$.  In practice, this is usually a regular series
of values $t_n = - n\; \delta t$ with an integration step $\delta t$.

To date HMC simulations have used either the past solution as a trial
solution, $\chi_{trial} = \chi(t_1)$, or a linear~\cite{gltrs,Gupta}
extrapolation, $\chi_{trial} = 2 \chi(t_1) - \chi(t_2)$.  It is
natural to try to improve the estimate for the trial value by using
higher order polynomial extrapolations.  For example, if one uses an
N-th order polynomial to fit N+1 past values, the coefficients are
given by
\begin{equation}
       c_k~=~ (-1)^{k-1}~ {N! \over k! (N-i)!} \; .
\label{coepoli}
\end{equation}
Although substantial improvements can be made using higher order
polynomial extrapolations, we found that the method breaks down for
large N. We have also considered using over constrained fits and some
estimators based on a Kalman filter. But we soon discovered that a
more appealing approach to the extrapolation problem.

Since the conjugate gradient method is in fact just a minimal residual
technique confined to the Krylov~\cite{Golub} subspace spanned by
vectors $A^{j-1}\chi_{trial}$, why not start by examining a
``smarter'' subspace based on past success for nearby times?  In this
spirit, we suggest determining the coefficients $c_n$ by minimizing
the functional,
\begin{equation}
        \Psi[\chi] =  
        \chi^\dagger M^\dagger M \chi - 
        \varphi^\dagger \chi -  \chi^\dagger \varphi \ ,
\end{equation}
which is the same functional minimized by the Conjugated Gradient
method itself.  This corresponds to the minimization of the norm of the
residual in the norm of the inverse matrix,
\begin{equation}
         r^\dagger {1 \over M^{\dagger} M } r   =  
        \chi^\dagger M^\dagger M \chi - 
        \varphi^\dagger \chi -  \chi^\dagger \varphi  
        + b^\dagger b \; ,
\end{equation}
in the subspace spanned by $\chi_i \equiv \chi(t_i)$, where $r = \phi
- M^\dagger M \chi$ and $b \equiv (M^\dagger)^{-1} \varphi$.
Therefore we call this method ``Chronological Inverter by Minimal
Residual Extrapolation'' or MRE.  The minimization condition reduces
to
\begin{equation}
  \sum_{j=1}^{N}~{ \chi_i }^\dagger M^\dagger M \chi_j \; c_j = 
   { \chi_i }^{\dagger} \varphi  \; .
\end{equation}
The only technical problem is that this system can be poorly
conditioned because the past solutions $ \chi_i $ differ from each
other by order $ \delta t $. However, if properly handled, this
instability should not affect the resulting minima of the quadratic
form in the span of the vectors, $ \chi_i $. We have explored other
methods of extrapolations to some extent, in particular other slightly
different definitions of residue (not in the norm of the inverse
matrix) were examined, but nothing appears at present to out-perform
the Minimal Residual Extrapolation method.

\vskip 0.7cm 

{\bf V. Chronological Inverter by Minimal Residual Extrapolation}

We implemented this method by calling a routine that minimizes the
residual in the subspace of the past solutions, right before each CG
in the usual leapfrog.  This routine has as inputs the set of past
vectors $\chi_i, ~ i=1...N $ and returns a vector $\chi_{trial}$ that
minimizes  $\Psi[\chi]$ in the {\it span}($\chi_i$):
\begin{center} {\bf MRE Algorithm} \end{center}
\begin{itemize}
\item Construct an orthogonal basis $v_i$ 
in the {\it span}($\chi_i$) using the Gram-Schmidt procedure.

\item  Form the sub-matrix $G_{n m} ={ v_n }^\dagger M^\dagger M v_m  $, 
and the vector $ b_n = { v_n }^{\dagger} \varphi$.

\item  Solve $G_{n m} a_m = b_n$ using the Gauss-Jordan
method.

\item Return the trial vector $ \chi_{trial} = \sum_{n=1}^{N}~
a_n v_n $.
\end{itemize}
Because $\chi_i$'s are nearly linearly dependent there is a numerical
instability, and the Gram-Schmidt procedure does not produce a true
basis of {\it span}($\chi_i$) due to round off error.  Hoverer {\it
span}($v_i$) is a subspace close enough to the relevant subspace of
the old solutions. Because the accumulated error is larger at the late
stages of the Gram-Schmitt iteration, the ordering of the $\chi_i$ is
important. We get better results if we number the $\chi_i$'s from the
newest to the oldest.  This way we quite accurately pick up the new
relevant directions, while the old ones, which are less relevant, are
computed with more round off error.  In this way we can handle the
numeric instability and turn it to our advantage.

This method requires $(N^2+5N)/2$ dot products and $N$ $M\chi$
matrix-vector applications, and the storage of $3N$ past
pseudo-fermion configurations.  There are other implementations of the
above concepts which can do the same thing using less memory. But most
of them suffer from numerical instabilities. The best we found, and
actually used in our production runs, is instead of storing $\chi's$,
store $v's$, keeping them in the right order - most recent first. This
reduces the memory requirement to the storage of $2 N$ pseudo-fermion
configurations, and has the same performance as the original method.
We checked this method using double precision, and there was no visible
improvement by reducing the round off-error.

In Table~\ref{tab:tavola1}, we record the mean number of CG steps
required to reach the solution normalized relative to starting with
$\chi_{trial}=\varphi$.  We used the stopping condition that the
normalized squared residual~(\ref{eq:stopping}) was smaller than $10^{-14}$.

\begin{figure}
\epsfxsize=15.5cm
\epsfysize=13.0cm
$$
\epsfbox{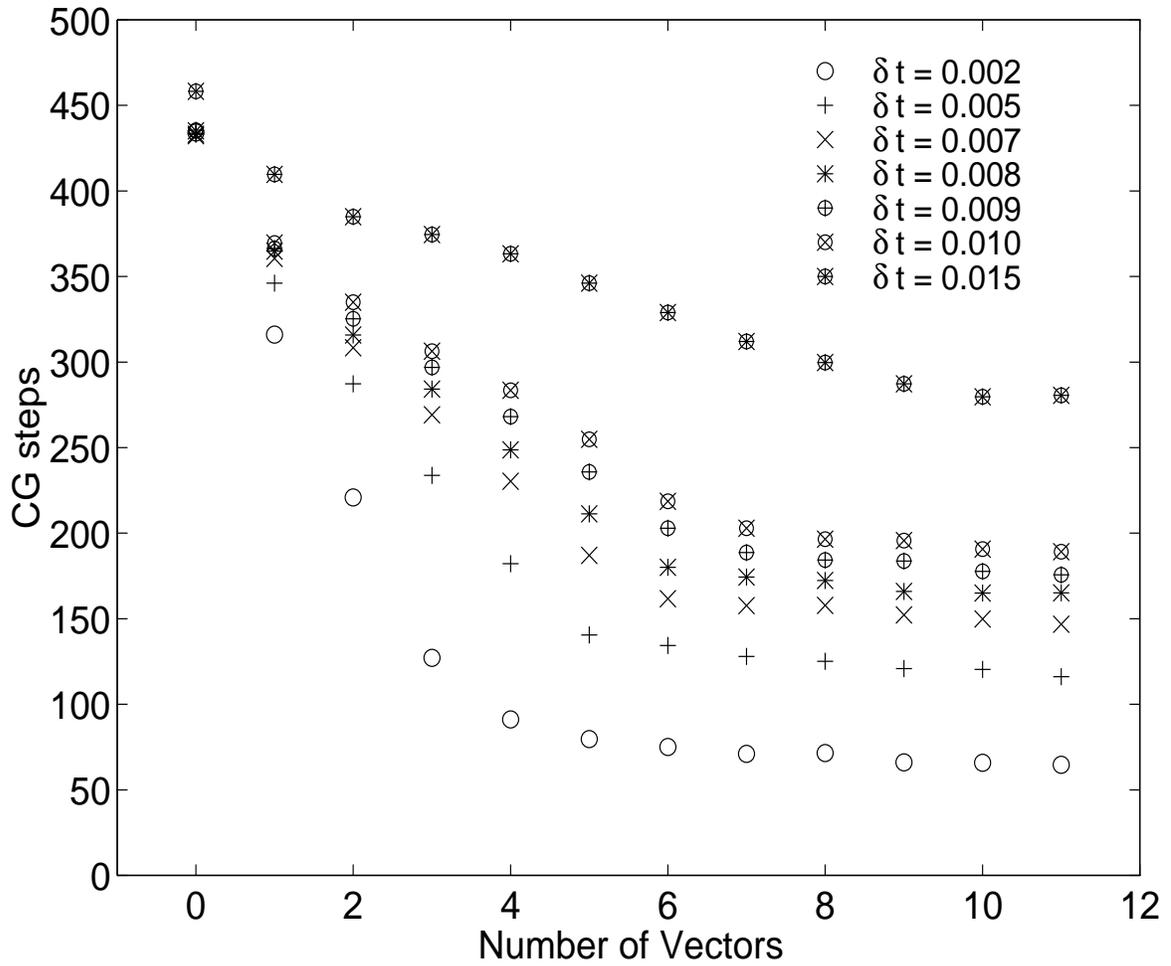} 
$$
\caption{Number of CG steps vs. number of vectors for various
values of \protect{$\delta t$} for the Minimal Residual Extrapolation method.
\label{fig:CGsteps}}
\end{figure}

\begin{figure}
\epsfxsize=15.5cm
\epsfysize=13.0cm
$$
\epsfbox{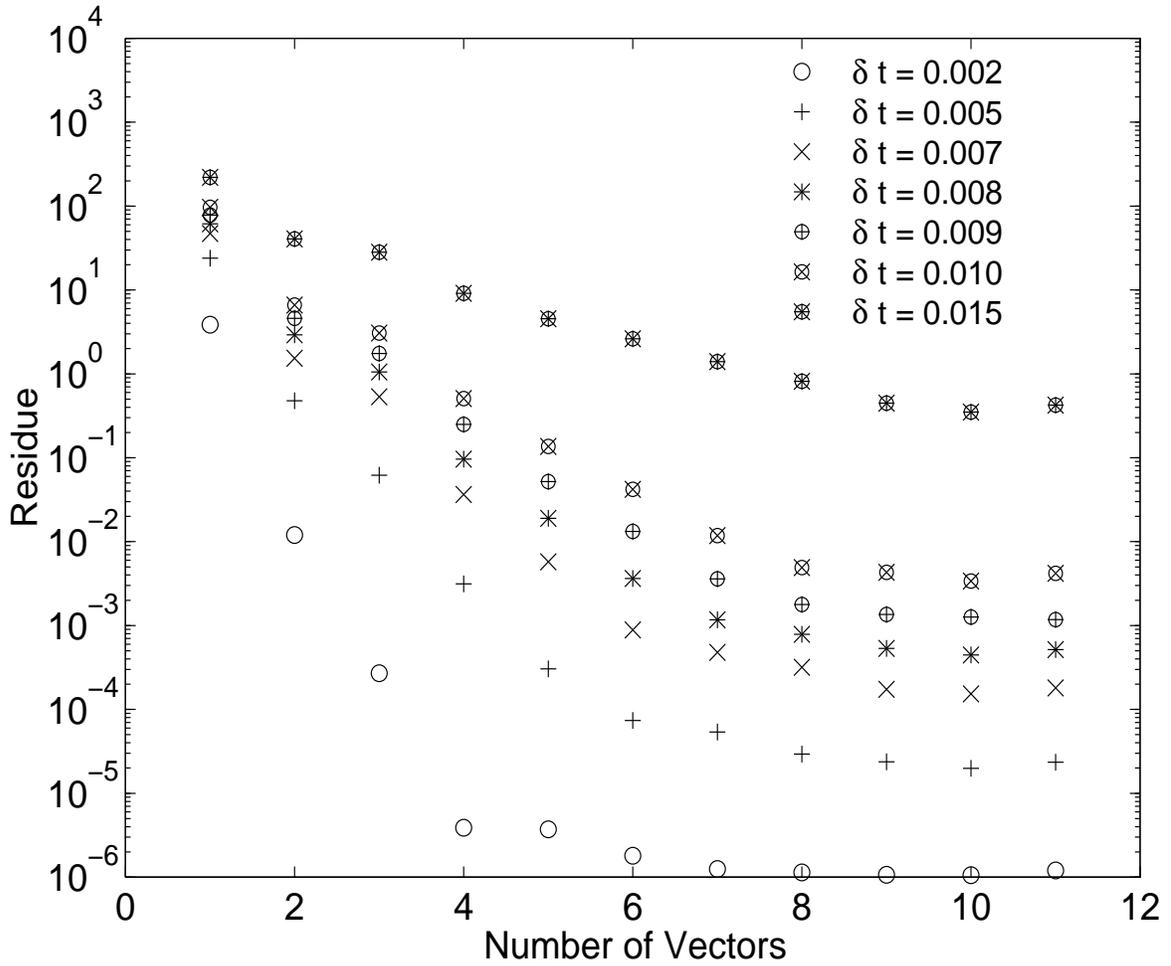} 
$$
\caption{Starting residue vs. number of vectors for various
values of \protect{$\delta t$} for the Minimal Residual Extrapolation method.
\label{fig:residue}}
\end{figure}

\begin{figure}
\epsfxsize=15.5cm
\epsfysize=13.0cm
$$
\epsfbox{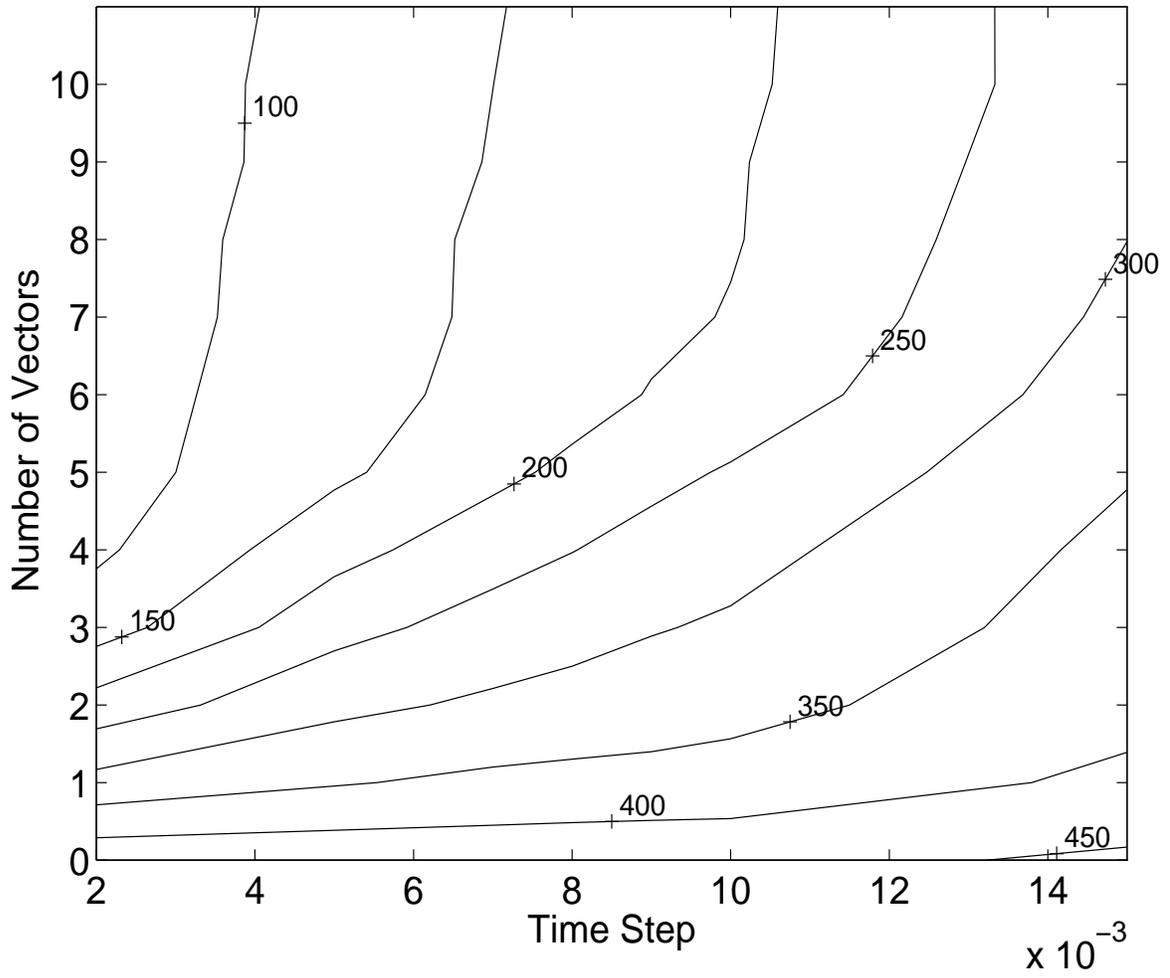} 
$$
\caption{ Contour plot of number of CG steps. \label{fig:CGcont}}
\end{figure}

The data (see Table~\ref{tab:tavola1} and Figures~\ref{fig:CGsteps},
~\ref{fig:residue} ), suggest that the more vectors you keep from the
past the better starting residue you achieve, and as a result the
fewer CG steps you need in order to converge to a given accuracy.
Furthermore the number of CG steps is decreasing as $\delta t$ is
decreasing.  We have also presented our results as a contour plot in
Figure~\ref{fig:CGcont} for the number of iterations as a function of
the number of vectors versus the time step.

\begin{table}[hbt]
\vskip 1cm 
\setlength{\tabcolsep}{1.2pc}
%\newlength{\digitwidth} \settowidth{\digitwidth}{\rm 0}
\catcode`?=\active \def?{\kern\digitwidth}
\begin{tabular}{|l|lllllll|}
\hline
N=0                 & 1.00  & 1.00  & 1.00  & 1.00  & 1.00  & 1.00  & 1.00  \\
N=1                 & 0.73  & 0.80  & 0.84  & 0.85  & 0.85  & 0.86  & 0.90  \\
N=2                 & 0.52  & 0.67  & 0.72  & 0.73  & 0.76  & 0.78  & 0.84  \\
N=3                 & 0.30  & 0.55  & 0.63  & 0.67  & 0.70  & 0.72  & 0.82  \\
N=4                 & 0.21  & 0.44  & 0.55  & 0.59  & 0.63  & 0.67  & 0.79  \\
N=5                 & 0.18  & 0.33  & 0.45  & 0.51  & 0.56  & 0.61  & 0.76  \\
N=6                 & 0.18  & 0.32  & 0.38  & 0.43  & 0.49  & 0.53  & 0.72  \\
N=7                 & 0.17  & 0.30  & 0.38  & 0.42  & 0.45  & 0.48  & 0.68  \\
N=8                 & 0.17  & 0.30  & 0.37  & 0.41  & 0.42  & 0.47  & 0.66  \\
N=9                 & 0.15  & 0.28  & 0.36  & 0.39  & 0.44  & 0.47  & 0.63  \\
N=10                & 0.15  & 0.29  & 0.36  & 0.39  & 0.42  & 0.45  & 0.61  \\
N=11                & 0.15  & 0.27  & 0.35  & 0.39  & 0.42  & 0.45  & 0.62  \\
\hline $\delta t$   & 0.002 & 0.005 & 0.007 & 0.008 & 0.009 & 0.010 & 0.015 \\
\hline
\end{tabular}
\caption{ \em
Number of CG steps needed to converge to the solution from minimum
residual extrapolation. The table is normalized with respect to N = 0
(i.e. no extrapolation, $\chi_{trial} = \varphi$).  The statistical
errors are of the order of 10\%.
\label{tab:tavola1}}
\vskip 1cm 
\end{table}

\begin{figure}

\epsfxsize=15.5cm
\epsfysize=13.0cm
$$
\epsfbox{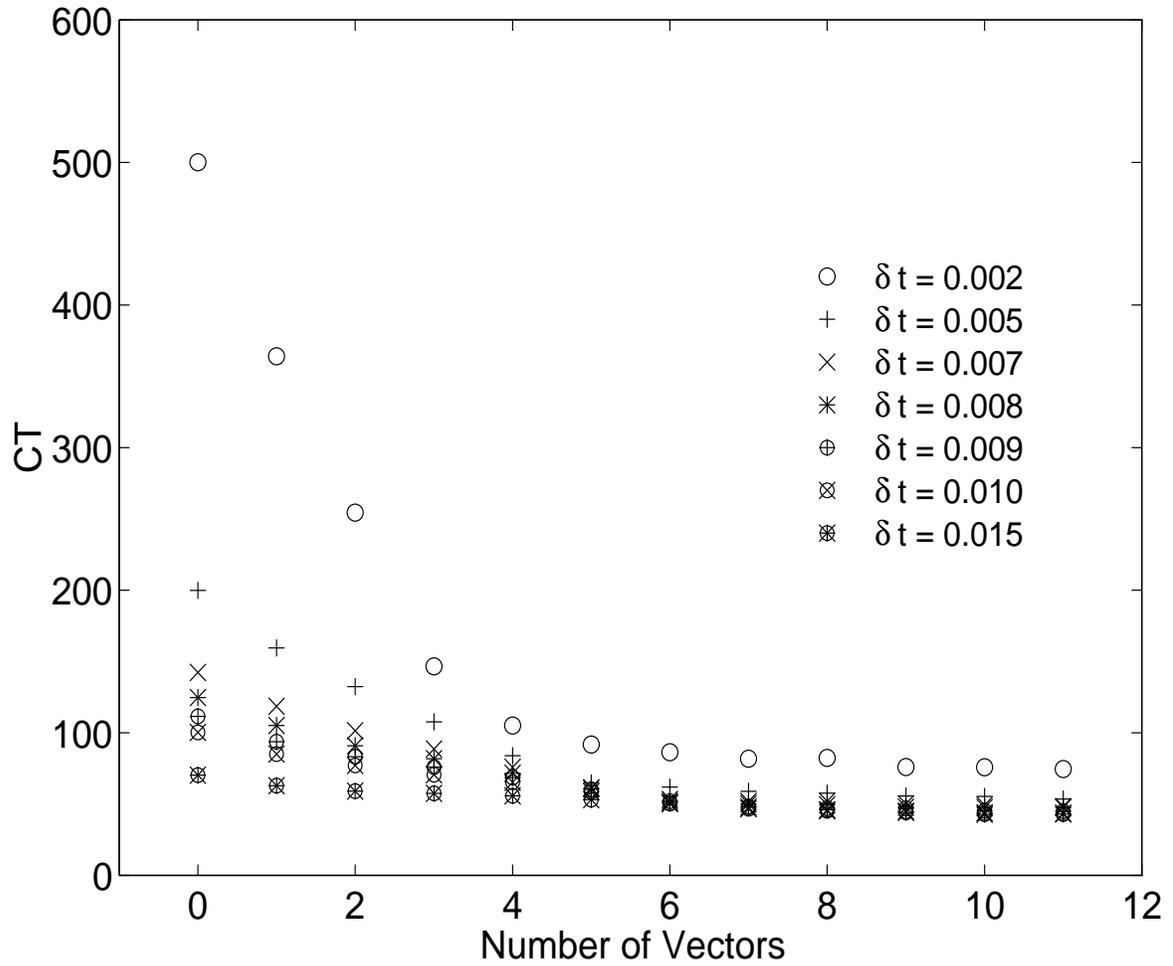} 
$$
\caption{ CT vs. number of vectors for various
values of \protect{$\delta t$} for the Minimal Residual Extrapolation method.
\label{fig:CE}}
\end{figure}

\begin{figure}
$$
\epsfxsize=15.5cm
\epsfysize=13.0cm
\epsfbox{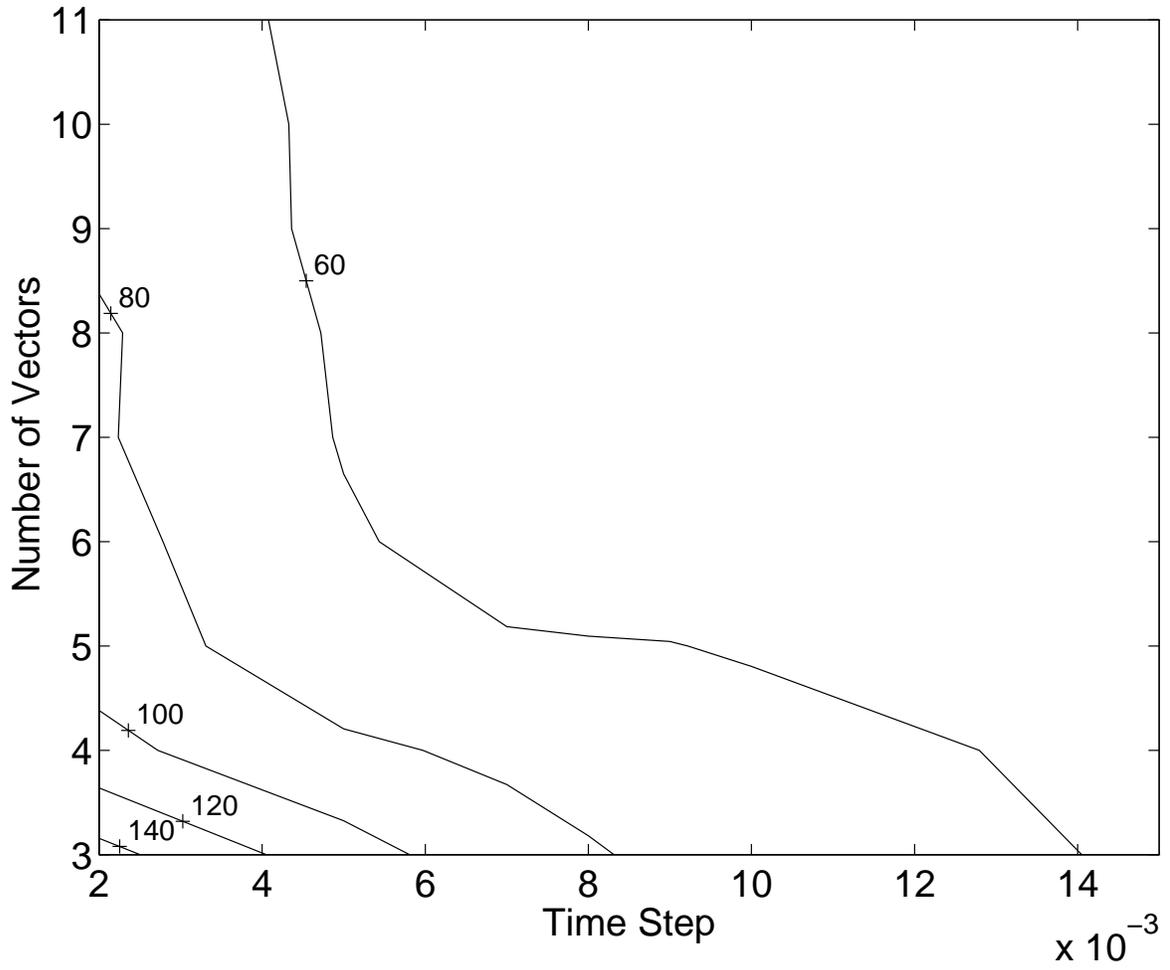} 
$$
\caption{ Contour plot of CT for the Minimal Residual Extrapolation method.
\label{fig:CEcont}}
\end{figure}

As explained above, the performance of the method given in
``computational time'' CT is summarized in Table~\ref{tab:tavola2},
and in Figures~\ref{fig:CE},\ref{fig:CEcont}.  It is interesting to
note how CT depends on the number of the vectors retained.  Unlike the
number of CG steps, CT does not decrease as $\delta t$ becomes
smaller.  Moreover, one can see from Figure~\ref{fig:CEcont} that
there is a range of $\delta t$ where the performance is relatively
insensitive to $\delta t$.  As you decrease the step size, you gained
back almost the same performance by reducing proportionally the number
of CG steps required for convergence in each step.  As discussed
earlier, smaller step sizes have the extra advantage of improving the
acceptance rate.

\begin{table}[hbt]
\vskip 1cm
\setlength{\tabcolsep}{1.2pc}
%\newlength{\digitwidth} \settowidth{\digitwidth}{\rm 0}
\catcode`?=\active \def?{\kern\digitwidth}
\begin{tabular}{|l|lllllll|}
\hline 
N=0                 & 5.86 & 2.34 & 1.66 & 1.46 & 1.30 & 1.17 & 0.78 \\
N=1                 & 4.29 & 1.88 & 1.40 & 1.24 & 1.10 & 1.00 & 0.70 \\
N=2                 & 3.03 & 1.57 & 1.20 & 1.07 & 0.98 & 0.91 & 0.66 \\
N=3                 & 1.78 & 1.29 & 1.06 & 0.96 & 0.91 & 0.84 & 0.64 \\
N=4                 & 1.23 & 1.02 & 0.92 & 0.86 & 0.82 & 0.78 & 0.62 \\
N=5                 & 1.08 & 0.78 & 0.74 & 0.74 & 0.73 & 0.71 & 0.59 \\
N=6                 & 1.04 & 0.74 & 0.64 & 0.63 & 0.63 & 0.62 & 0.56 \\
N=7                 & 0.99 & 0.71 & 0.63 & 0.61 & 0.58 & 0.57 & 0.53 \\
N=8                 & 0.98 & 0.69 & 0.62 & 0.60 & 0.57 & 0.55 & 0.51 \\
N=9                 & 0.88 & 0.67 & 0.60 & 0.58 & 0.57 & 0.54 & 0.49 \\
N=10                & 0.90 & 0.67 & 0.60 & 0.57 & 0.55 & 0.53 & 0.48 \\
N=11                & 0.88 & 0.64 & 0.59 & 0.57 & 0.54 & 0.53 & 0.48 \\ 
\hline $\delta t$  &  0.002 & 0.005 & 0.007 & 0.008 & 0.009 & 0.010 & 0.015 \\ \hline
\end{tabular}
\caption{\em
Computational Time (CT) starting from minimum residual extrapolation.
The table is normalized to 1 at $\delta t=0.010$
for N = 1 ( $\chi_{trial}=\chi_{last}$). 
\label{tab:tavola2}}
\vskip 1cm
\end{table}

\pagebreak

\vskip 0.7cm 

{\bf VI. Comparison with the Polynomial Extrapolation Method}

The polynomial extrapolation method has the advantage that it requires
very little computational effort, just a local sum on each lattice
point with fixed coefficients given once and for all by equation
(\ref{coepoli}), costing less than a single CG step.  For a polynomial
of order $N$, the only storage requirement is for the previous $N$
pseudo-fermion configurations, $\chi(t_i)$.

\begin{figure}
\epsfxsize=15.5cm
\epsfysize=13.0cm
$$
\epsfbox{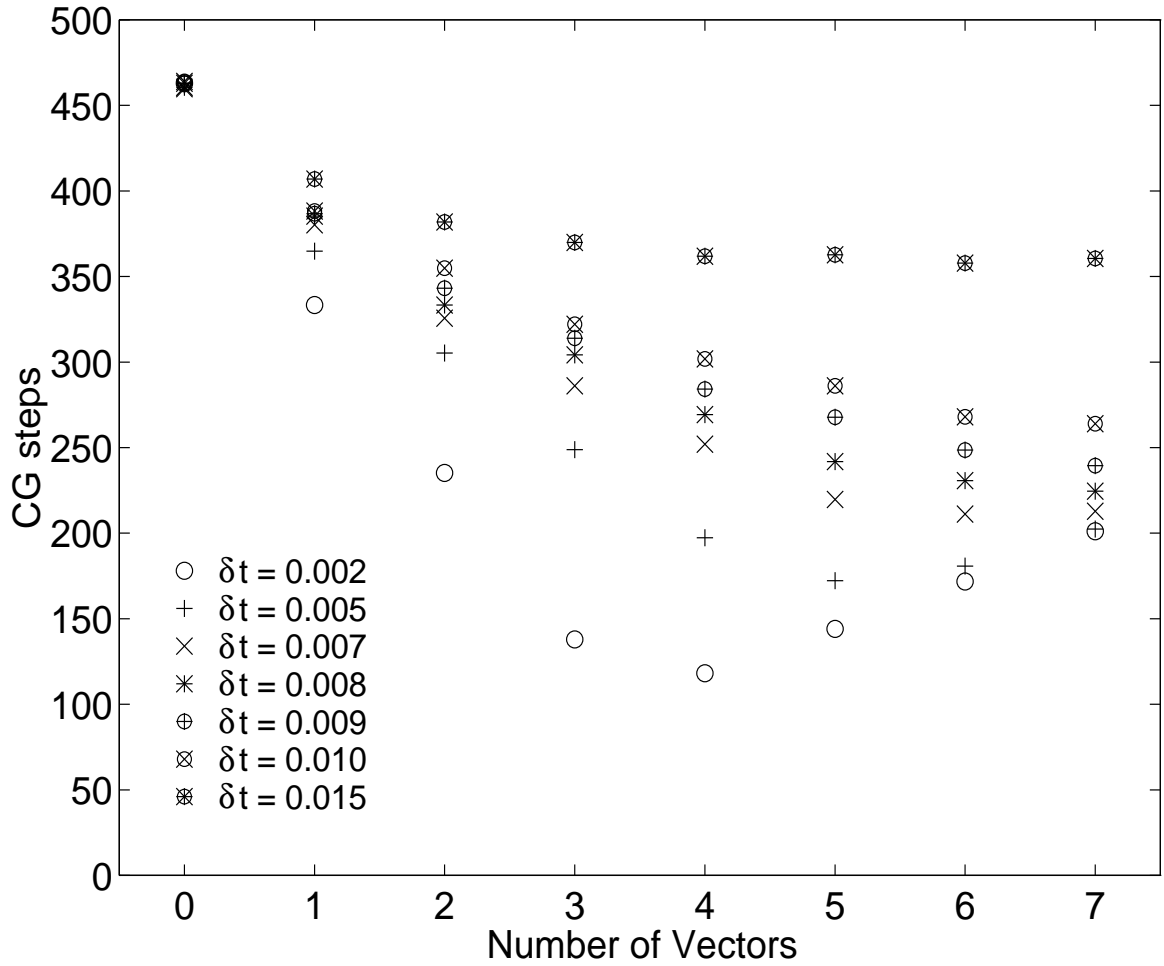} 
$$
\caption{Number of CG steps vs. number of vectors for various
values of \protect{$\delta t$} for the Polynomial extrapolation.
 \label{fig:CGsteps_p}}
\end{figure}

\begin{figure}
\epsfxsize=15.5cm
\epsfysize=13.0cm
$$
\epsfbox{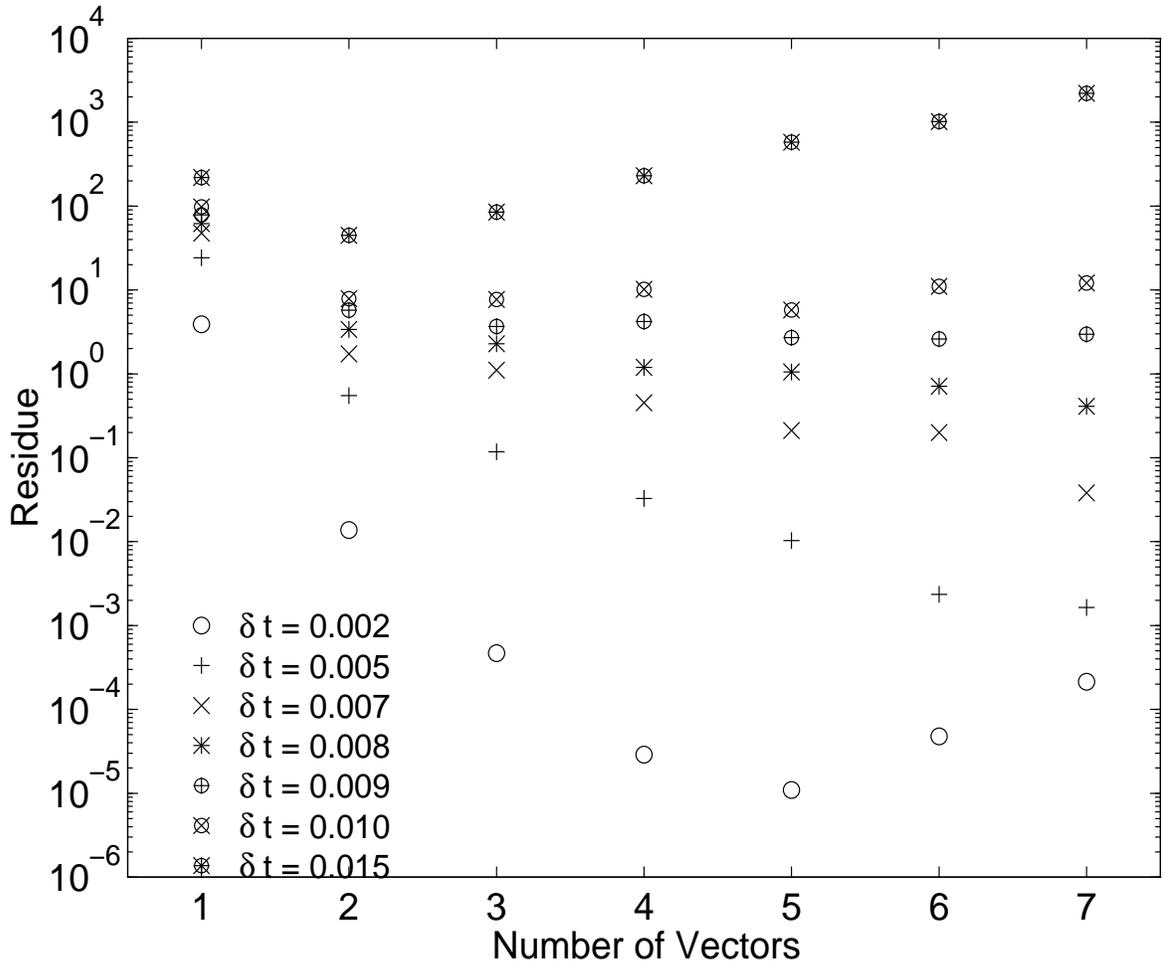} 
$$
\caption{Starting residue vs. number of vectors for various
values of \protect{$\delta t$} for the polynomial extrapolation.
\label{fig:residue_p}}
\end{figure}

\begin{figure}
\epsfxsize=15.5cm
\epsfysize=13.0cm
$$
\epsfbox{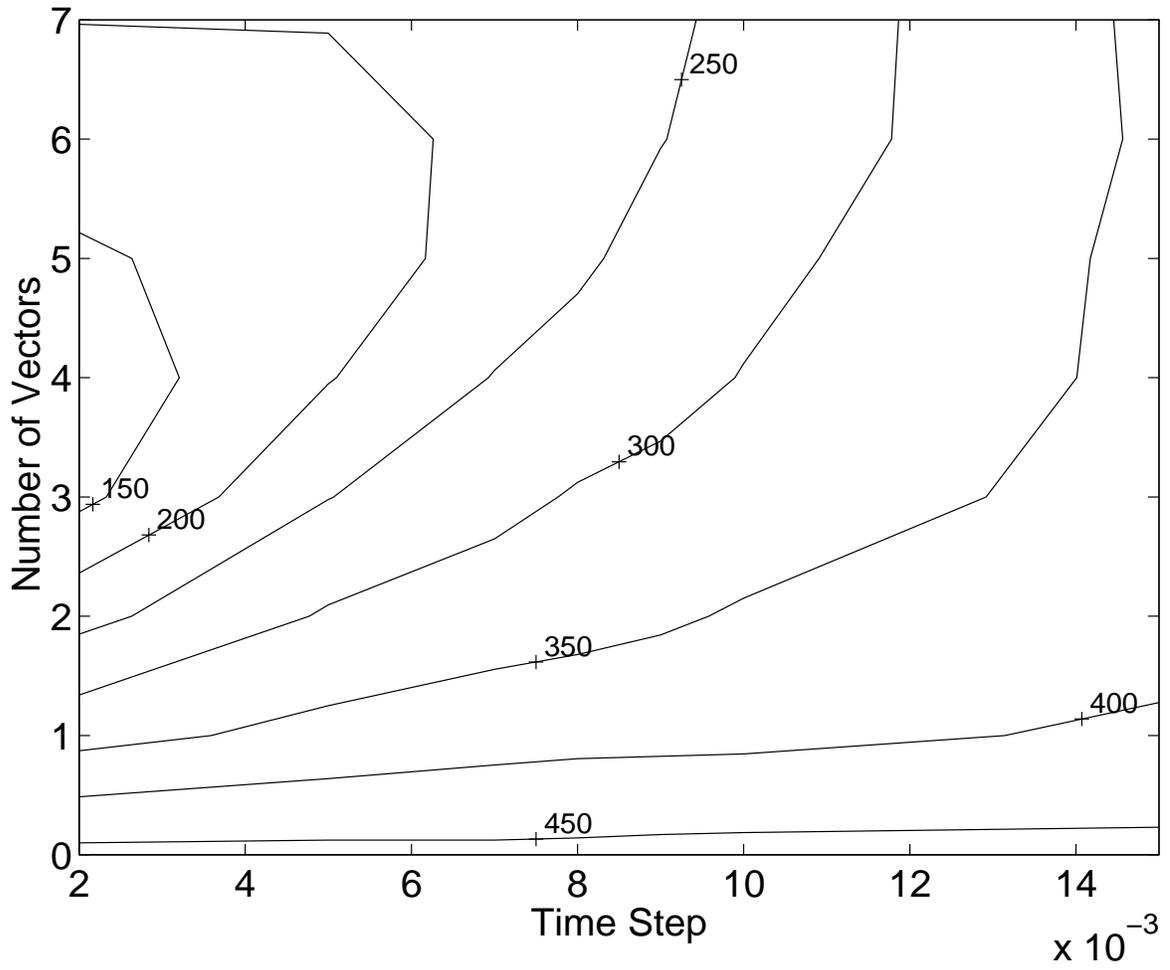} 
$$
\caption{ Contour plot of number of CG steps for the Polynomial extrapolation.
\label{fig:CGcont_p}}
\end{figure}

\begin{figure}

\epsfxsize=15.5cm
\epsfysize=13.0cm
$$
\epsfbox{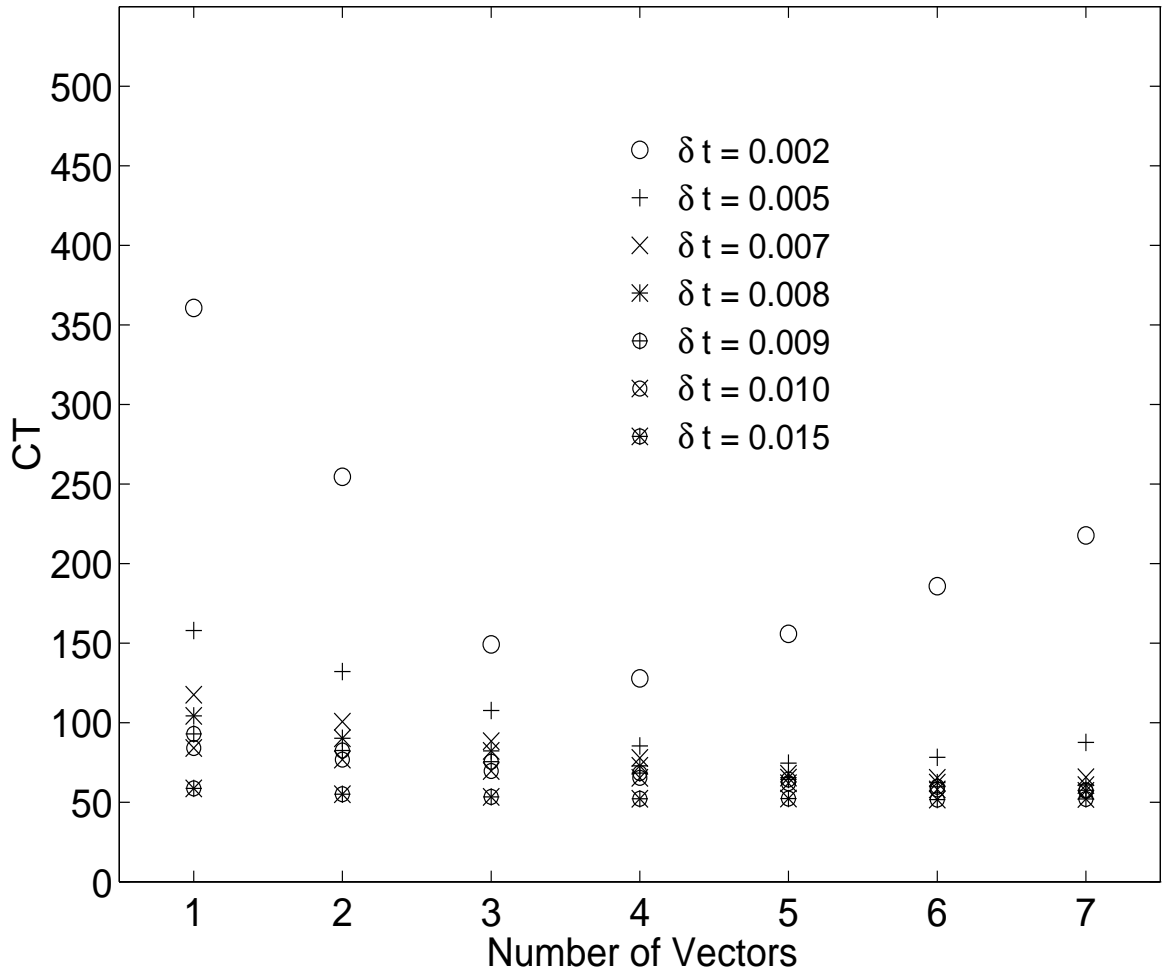} 
$$
\caption{ CT vs. Number of vectors for various
values of \protect{$\delta t$} for the Polynomial extrapolation. 
\label{fig:CE_p}}
\end{figure}

\begin{figure}
$$
\epsfxsize=15.5cm
\epsfysize=13.0cm
\epsfbox{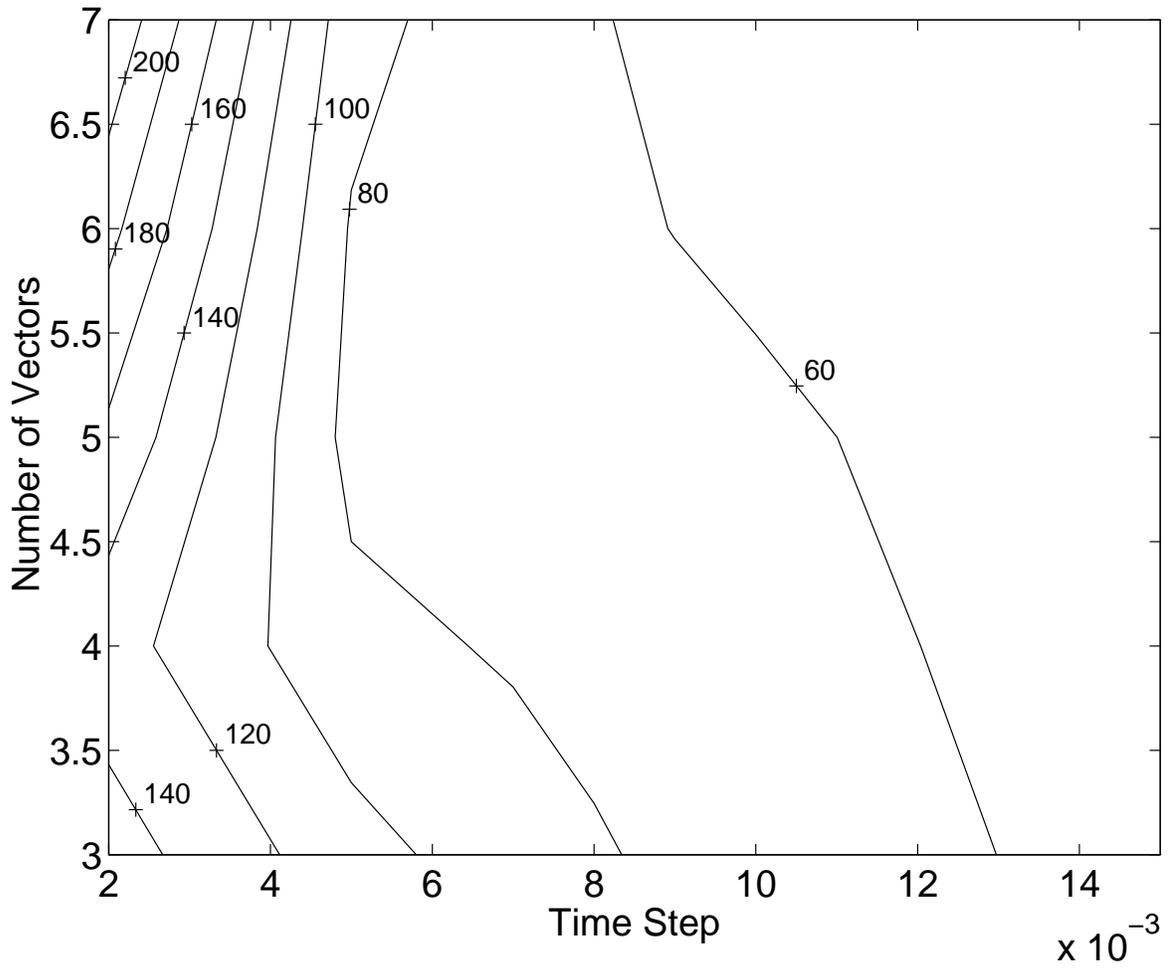} 
$$
\caption{ Contour plot of CT for the Polynomial extrapolation.
\label{fig:CEcont_p}}
\end{figure}

Results of the polynomial extrapolation are shown in
Tables~\ref{tab:tavola3},\ref{tab:tavola4} and Figures~\ref
{fig:CGsteps_p},\ref{fig:residue_p},\ref{fig:CGcont_p},\ref{fig:CE_p},
\ref{fig:CEcont_p}. Again, to compare
efficiencies, the CG iterations should be divided by $\delta t$, so we
compare the total number of CG steps needed to evolve the system for a
fixed distance in configuration space.  Note that if $\delta t$ is
large, the overall performance is good, but the acceptance will drop
drastically.  If $\delta t$ is small, the extrapolation is excellent,
but the system will evolve too slowly in phase space.  It is
noteworthy that we find a window in $\delta t$ where both the
acceptance is good and the extrapolation works well.

\begin{table}[hbt]
\vskip 1cm
\setlength{\tabcolsep}{1.2pc}
%newlength{\digitwidth} \settowidth{\digitwidth}{\rm 0}
\catcode`?=\active \def?{\kern\digitwidth}
\begin{tabular}{|l|lllllll|}
\hline 
N=0                 & 1.00 & 1.00 & 1.00 & 1.00 & 1.00 & 1.00 & 1.00\\
N=1                 & 0.72 & 0.79 & 0.82 & 0.83 & 0.84 & 0.84 & 0.88\\
N=2                 & 0.51 & 0.66 & 0.70 & 0.72 & 0.74 & 0.77 & 0.83\\
N=3                 & 0.30 & 0.54 & 0.62 & 0.66 & 0.68 & 0.70 & 0.80\\
N=4                 & 0.26 & 0.43 & 0.55 & 0.58 & 0.62 & 0.65 & 0.78\\
N=5                 & 0.31 & 0.37 & 0.48 & 0.52 & 0.58 & 0.62 & 0.79\\
N=6                 & 0.37 & 0.39 & 0.46 & 0.50 & 0.54 & 0.58 & 0.77\\
N=7                 & 0.44 & 0.44 & 0.46 & 0.49 & 0.52 & 0.57 & 0.78\\
\hline $\delta t$  &  0.002 & 0.005 & 0.007 & 0.008 & 0.009 & 0.010 & 0.015 \\ \hline
\end{tabular}
\caption{\em Number of CG steps to reach the solution 
for polynomial extrapolation.  Normalized to 1 for N = 0 ($\chi_{trial}=\varphi$). 
\label{tab:tavola3}}
\vskip 1cm
\end{table}

\begin{table}[hbt]
\vskip 1cm 
\setlength{\tabcolsep}{1.2pc}
%newlength{\digitwidth} \settowidth{\digitwidth}{\rm 0}
\catcode`?=\active \def?{\kern\digitwidth}
\begin{tabular}{|l|lllllll|}
\hline 
N=0                 & 5.96 & 2.38 & 1.69 & 1.48 & 1.32 & 1.20 & 0.79 \\
N=1                 & 4.29 & 1.88 & 1.40 & 1.24 & 1.11 & 1.00 & 0.70 \\
N=2                 & 3.03 & 1.57 & 1.20 & 1.07 & 0.98 & 0.91 & 0.66 \\
N=3                 & 1.78 & 1.28 & 1.05 & 0.98 & 0.90 & 0.83 & 0.64 \\
N=4                 & 1.52 & 1.01 & 0.93 & 0.87 & 0.81 & 0.78 & 0.62 \\
N=5                 & 1.85 & 0.89 & 0.81 & 0.78 & 0.77 & 0.74 & 0.62 \\
N=6                 & 2.21 & 0.93 & 0.78 & 0.74 & 0.71 & 0.69 & 0.61 \\
N=7                 & 2.59 & 1.04 & 0.78 & 0.72 & 0.69 & 0.68 & 0.62 \\
\hline $\delta t$  &  0.002 & 0.005 & 0.007 & 0.008 & 0.009 & 0.010 & 0.015 \\ \hline
\end{tabular}
\caption{\em Computational Time (CT) starting from polynomial extrapolation.
The table is normalized to 1 at
$\delta t=0.010$   for N = 1 ($\chi_{trial}=\chi_{last}$).
\label{tab:tavola4}}
\vskip 1cm 
\end{table}

We also note that the coefficients of the Minimal Residual Extrapolation method
(\ref{timecoef}), which \'a priori are generic complex numbers, are in
fact very close to coefficients in the polynomial extrapolation
(\ref{coepoli}) for the first few orders.  One way to understand this
coincidence is to observe that for a smooth evolution the determination
of coefficients $c_i$,
\begin{equation}
       \chi_{trial} = c_1 \; \chi(t_1) + c_2 \; \chi(t_2) 
        + \cdots + c_N \; \chi(t_N),
\end{equation}
by a polynomial fit is {\bf equivalent} to  fixing the coefficients by making
a Taylor expansion of each term, $\chi(t_n)$, in $t_n = n \delta t$
canceling all contributions to $O(\delta t^N)$. To prove this we
solve the constraints
\begin{equation}
      \sum_{n=1}^{N} (t_i)^{n-1} \; y_n = \chi(t_i) \; ,
\label{equ:poly}
\end{equation}
for a polynomial fit $y(t)~=~y_1 + t \; y_2 +\cdots + \;t^{N-1}\; y_N$
to find $y_1 = \chi_{trial}$. Then show that $y_1 = \sum_i c_i \chi(t_i)$
when we enforce the Taylor series constraints,
\begin{equation}
         \sum_{i=1}^{N} (t_i)^{n-1} \; c_i = \delta_{1,n} \ .
\label{equ:taylor}
\end{equation}

 From this exercise, we conclude that the success of the polynomial fit
probably results from the local convergence of the power series MD in
time.  On the other hand, if we compare the magnitude of the estimated
residual of the Minimal Residual Extrapolation (MRE) versus the
Polynomial Extrapolation (see
Figures~\ref{fig:residue}, \ref{fig:residue_p}), we notice that while
the polynomial actually deteriorates at high order, the minimal
residual continues to improve.  In our simulation for a polynomial
extrapolation of order of $N=5-6$, the deterioration becomes
appreciable.

In conclusion the increased efficiency of the MRE method is worth the
extra computational effort (equivalent to $N/2$ CG steps).  With a
Polynomial Extrapolation, one gets poor results on a few exceptional
lattices, whereas the MRE method is a more robust estimator, which
implements a kind of self-tuned extrapolation, never doing worse than
a polynomial fit of the same order.

\vskip 0.7cm

{\bf VII.  Projective Conjugate Gradient}

It is natural at this point to try to understand why the Chronological
Inverter is working and to explore more effective methods of using the
past information.  We have performed a large number of ``numerical
experiments'' to gain some understanding. In particular we have shown
that if one assumes complete knowledge of the inverse of the matrix in
the immediate past $A_1 \equiv A(t_1)$ and performs a perturbation
expansion in the difference $V = A - A_1$ convergence is poor unless
$\delta t$ is very small on the order of $10^{-3}$.  On the other hand,
although we know no way to efficiently implement the idea, if we use
$A_1$ as a preconditioner for the conjugate gradient iterations of A,
a spectacular acceleration of the conjugate gradient is achieved
 --- for $\delta t = 0.01$ converging to 
$R = 10^{-16}$ in about 15 iterations. In this way,
we are now convinced that the use of past information as a
preconditioner holds out even greater promise. We are engaged in
on going research in this regard. However, here we report a minor
extension of the current method along these lines.
 
\begin{figure}
\epsfxsize=15.5cm
\epsfysize=13.0cm
$$
\epsfbox{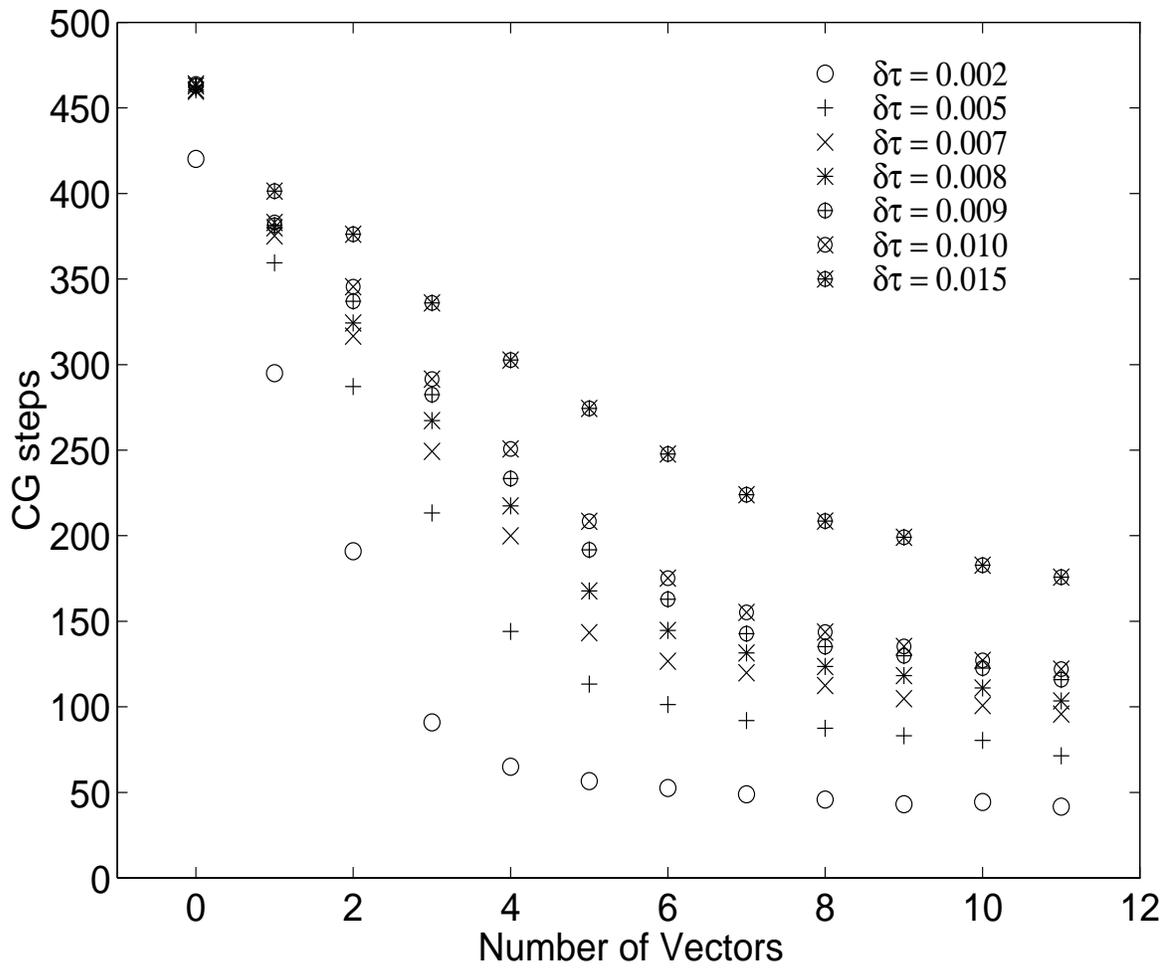} 
$$
\caption{Number of CG steps vs. number of vectors for various
values of \protect{$\delta t$} for the Projective CG.
 \label{fig:CGsteps_j}}
\end{figure}

\begin{figure}
\epsfxsize=15.5cm
\epsfysize=13.0cm
$$
\epsfbox{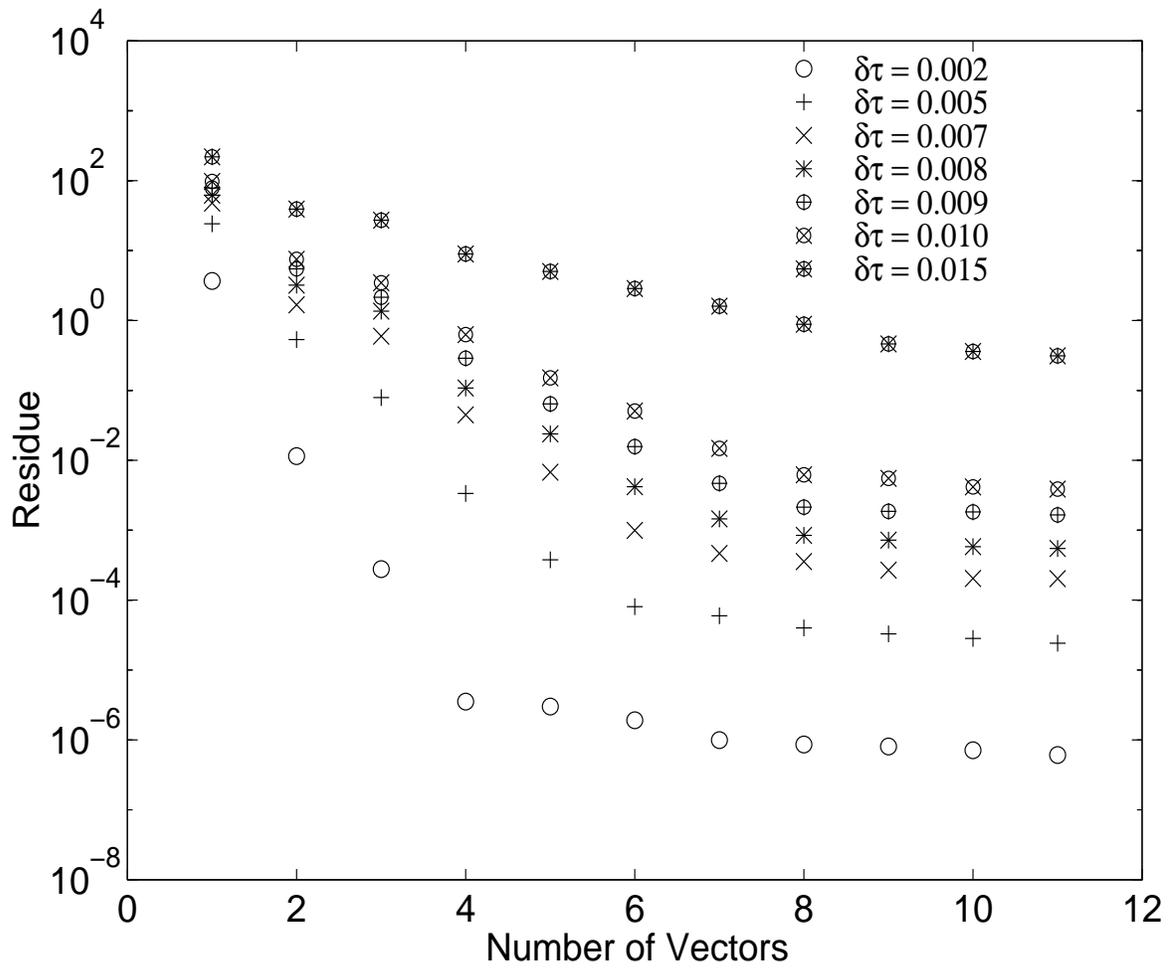} 
$$
\caption{Starting residue vs. number of vectors for various
values of \protect{$\delta t$} for the Projective CG.
\label{fig:residue_j}}
\end{figure}

\begin{figure}
\epsfxsize=15.5cm
\epsfysize=13.0cm
$$
\epsfbox{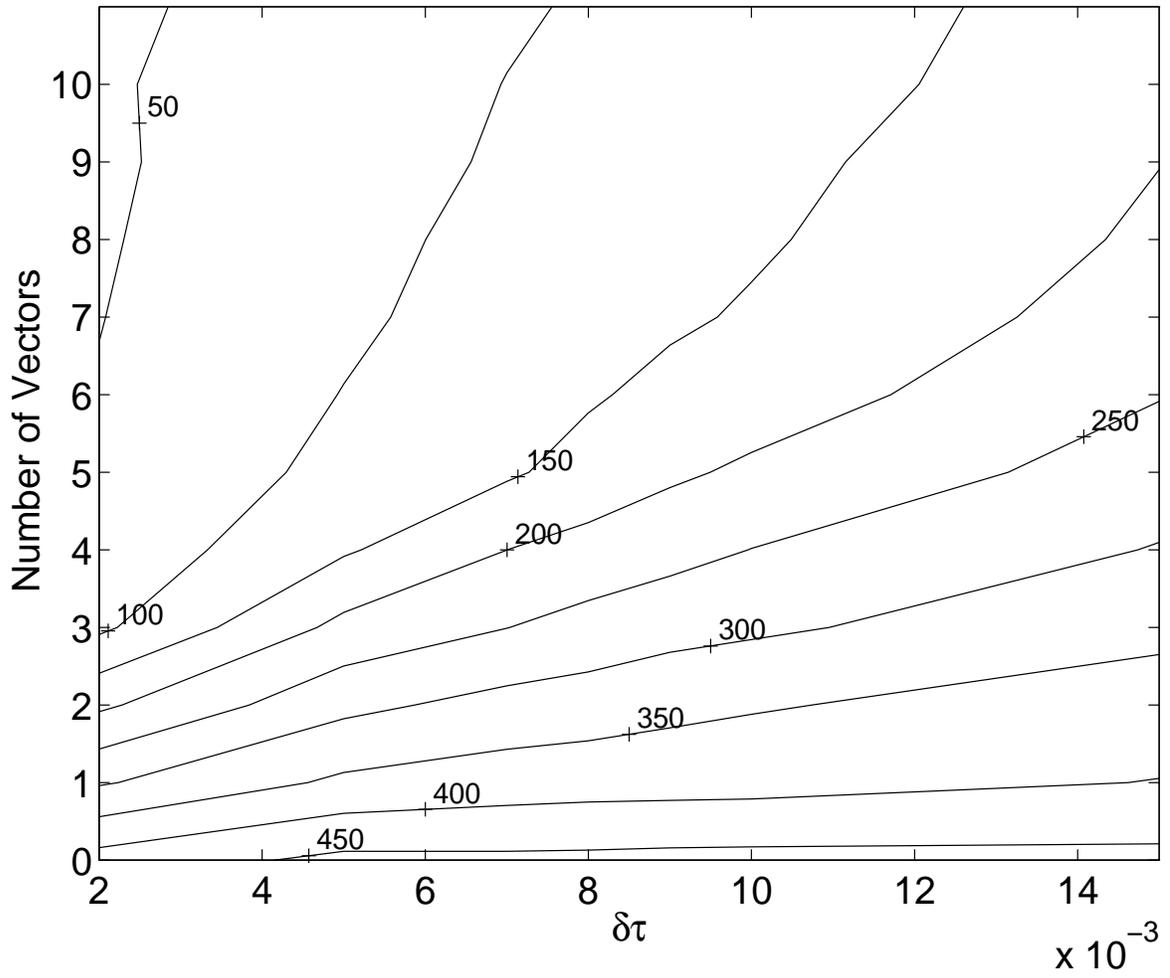} 
$$
\caption{ Contour plot of number of CG steps for the Projective CG.
\label{fig:CGcont_j}}
\end{figure}

\begin{figure}

\epsfxsize=15.5cm
\epsfysize=13.0cm
$$
\epsfbox{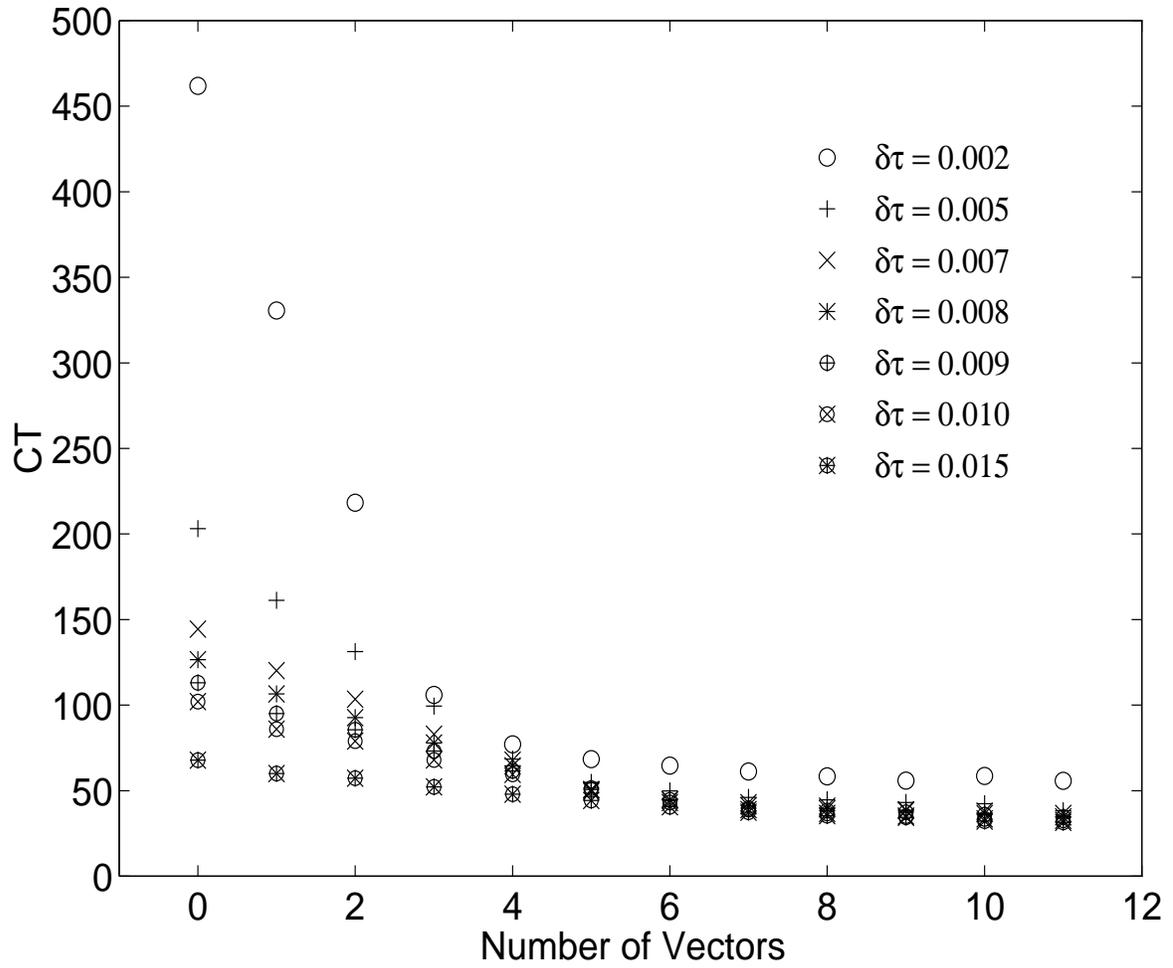} 
$$
\caption{ CT vs. Number of vectors for various
values of \protect{$\delta t$} for the Projective CG.
\label{fig:CE_j}}
\end{figure}

\begin{figure}
$$
\epsfxsize=15.5cm
\epsfysize=13.0cm
\epsfbox{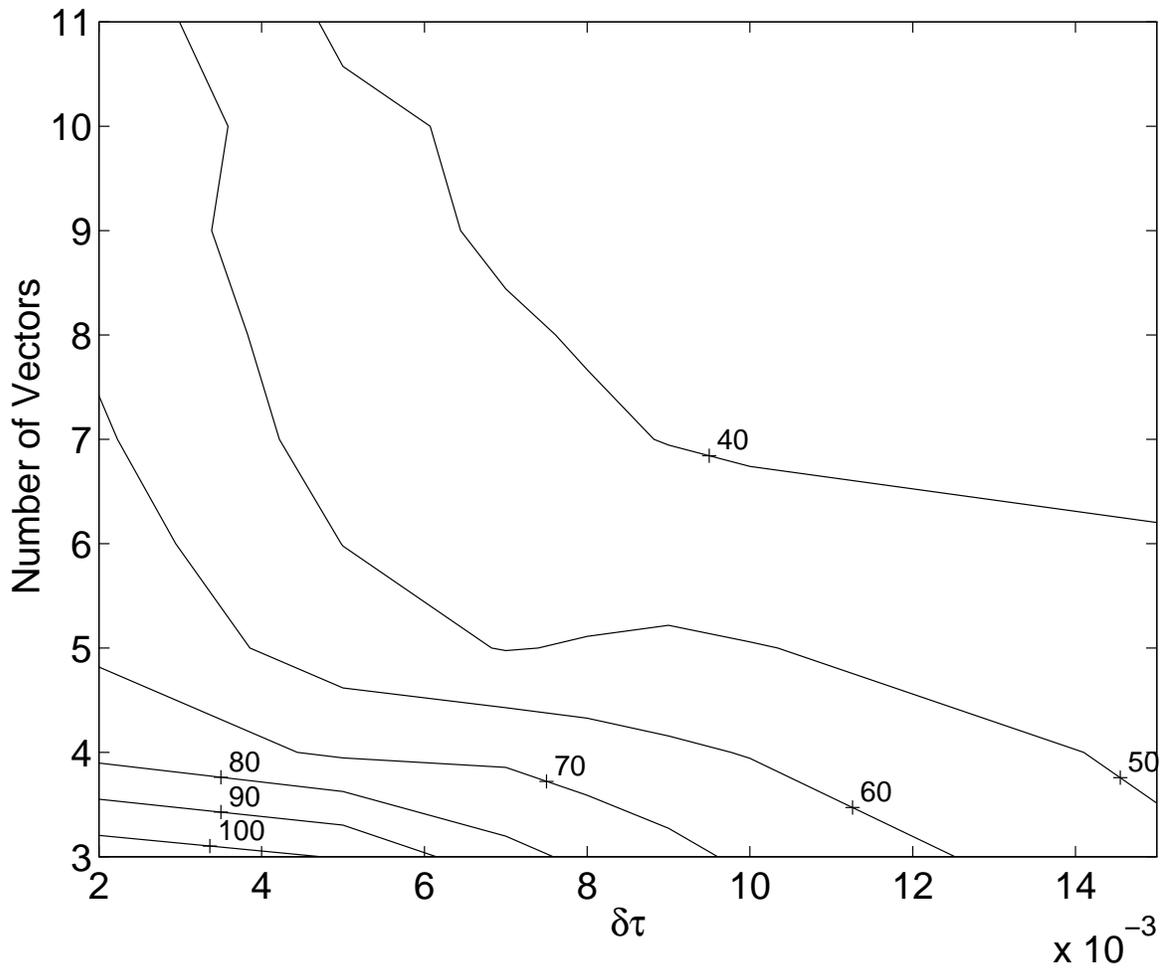} 
$$
\caption{ Contour plot of CT for the Projective CG.
\label{fig:CEcont_j}}
\end{figure}

In the Minimal Residual Extrapolation, we suggest re-using our past
solutions $\{\chi_i\}$ to construct a preconditioner for the
subsequent CG iterations.  The straight forward modification is as
follows.  Since we have already minimized the functional $\Psi(\chi)$
in the subspace spanned by $\chi_i$'s, it is reasonable to require
that all the subsequent search directions we generate in CG are also
$A$-conjugate to this subspace. This will ensure that at any step $K$
of CG, we obtain the global minimum of $\Psi(\chi)$ in the space
spanned by $\chi_i$'s and $p_k$ ($k =1..K$).  This requirement is
implemented by almost the same technique used in conventional
preconditioners for CG~\cite{Golub}. In each CG iteration, the
residual $r_k$ is replaced by an improved search direction vector
$z_k$.  Thus we propose for our Projected Conjugate Gradient (PCG)
method that the vector $z$ is chosen to be the A orthogonal component
of $r$ with respect to {\it span}($\chi_i$) or in terms of the basis
vectors basis $v_n$ defined above for our MRE algorithm (see Section
V)
\be
z = r - \lambda_n v_n,
\ee
where the condition $v^\dagger_n A z = 0$ means
\be
G_{n m}\lambda_m = v^\dagger_n M^\dagger M r
\ee
with $A = M^\dagger M$ and $G_{n m} = { v_n }^\dagger M^\dagger M v_m$
defined as before.  The coefficients $\lambda_n$ are found by solving
a small linear system using the Gauss-Jordan elimination method or the
LU factorization. The amount of computation for solving such a system
is very small, especially if one uses the LU factorization, since the
matrix $G_{n m}$ is computed once and the LU factors can be used in
subsequent steps.  As a result we modify the conjugate gradient
routine as follows:
\begin{center} {\bf PCG Algorithm} \end{center}
\begin{itemize}
\item Project residue: $z_k = 
r_k -  v_n (G^{-1})_{n,m} (v^\dagger_m M^\dagger M r_k)$  

\item Compute search direction:
       $p_{k}= z_{k-1} - \beta_{k-1} p_{k-1}$,
      where  $ \beta_{k-1} = r_{k-1}^\dagger z_{k-1}/ r_{k-2}^\dagger z_{k-2}$.
\item Compute new  solution:
	$\chi_k = \chi_{k-1} + \alpha_k p_k$, where 
        $ \alpha_k = r_{k-1}^\dagger z_{k-1}/p_k^{\dagger} M^\dagger M p_k$

\item Compute new residual:
	$r_k = r_{k-1} -   M^\dagger M  p_k$.
\end{itemize}
It can easily be shown that the above algorithm produces a set of
search vectors that are A-conjugate among themselves and to the
$v_i$'s.  The only time consuming step, needed every CG iteration, is
the computation of the right-hand side $ { v_n }^{\dagger} M^\dagger M
r_{k} $, but the vectors $\tilde v_n = M^\dagger M v_n$ have in fact
already been computed in the MRE step so this only involves N scalar
products.  For our implementation on the MIT 128-node CM-5, one
$M^\dagger M \chi$ operation is equivalent to 55 $ {\tilde v
}^{\dagger} r $ operations so with O(10) vectors the overhead is about
20\%.  This cost has been taken into account when analyzing the
performance of this algorithm. The results of our performance tests
are summarized in Tables~\ref{tab:tavola5},\ref{tab:tavola6} and in
Figures~\ref{fig:CGsteps_j},\ref{fig:residue_j},\ref{fig:CGcont_j},\ref
{fig:CE_j},\ref{fig:CEcont_j}. 

\begin{table}[hbt]
\vskip 1cm 
\setlength{\tabcolsep}{1.2pc}
%\newlength{\digitwidth} \settowidth{\digitwidth}{\rm 0}
\catcode`?=\active \def?{\kern\digitwidth}
\begin{tabular}{|l|lllllll|}
\hline
N=0                 & 1.00  & 1.00  & 1.00  & 1.00  & 1.00  & 1.00  & 1.00  \\
N=1                 & 0.65  & 0.79  & 0.82  & 0.84  & 0.84  & 0.84  & 0.88  \\
N=2                 & 0.42  & 0.63  & 0.70  & 0.71  & 0.74  & 0.76  & 0.83  \\
N=3                 & 0.20  & 0.47  & 0.55  & 0.59  & 0.62  & 0.64  & 0.74  \\
N=4                 & 0.14  & 0.32  & 0.44  & 0.48  & 0.51  & 0.55  & 0.67  \\
N=5                 & 0.12  & 0.25  & 0.32  & 0.37  & 0.42  & 0.46  & 0.60  \\
N=6                 & 0.12  & 0.22  & 0.28  & 0.32  & 0.36  & 0.39  & 0.54  \\
N=7                 & 0.11  & 0.20  & 0.26  & 0.29  & 0.31  & 0.34  & 0.49  \\
N=8                 & 0.10  & 0.19  & 0.25  & 0.27  & 0.30  & 0.32  & 0.46  \\
N=9                 & 0.09  & 0.18  & 0.23  & 0.26  & 0.29  & 0.30  & 0.44  \\
N=10                & 0.09  & 0.18  & 0.22  & 0.24  & 0.27  & 0.28  & 0.40  \\
N=11                & 0.09  & 0.16  & 0.21  & 0.23  & 0.25  & 0.27  & 0.39  \\
\hline $\delta t$   & 0.002 & 0.005 & 0.007 & 0.008 & 0.009 & 0.010 & 0.015 \\
\hline
\end{tabular}
\caption{ \em
Number of the projective CG steps needed to converge to the solution.
We used Minimal Residual Extrapolation to obtain the initial trial solution
$\chi$.  The table is normalized with respect to N = 0 (i.e. no
extrapolation, $\chi_{trial} = \varphi$).  The statistical errors are
of the order of 10\%.
\label{tab:tavola5}}
\vskip 1cm 
\end{table}

\begin{table}[hbt]
\vskip 1cm
\setlength{\tabcolsep}{1.2pc}
%\newlength{\digitwidth} \settowidth{\digitwidth}{\rm 0}
\catcode`?=\active \def?{\kern\digitwidth}
\begin{tabular}{|l|lllllll|}
\hline 
N=0                 & 5.38 & 2.36 & 1.68 & 1.47 & 1.32 & 1.18 & 0.79 \\
N=1                 & 3.85 & 1.88 & 1.40 & 1.24 & 1.11 & 1.00 & 0.70 \\ 
N=2                 & 2.54 & 1.53 & 1.20 & 1.08 & 1.00 & 0.92 & 0.67 \\
N=3                 & 1.23 & 1.16 & 0.97 & 0.91 & 0.85 & 0.79 & 0.61 \\
N=4                 & 0.90 & 0.80 & 0.79 & 0.75 & 0.72 & 0.69 & 0.55 \\
N=5                 & 0.80 & 0.64 & 0.58 & 0.59 & 0.60 & 0.59 & 0.51 \\
N=6                 & 0.75 & 0.58 & 0.52 & 0.52 & 0.52 & 0.50 & 0.47 \\
N=7                 & 0.71 & 0.54 & 0.50 & 0.48 & 0.46 & 0.45 & 0.44 \\
N=8                 & 0.68 & 0.52 & 0.48 & 0.46 & 0.45 & 0.43 & 0.41  \\
N=9                 & 0.65 & 0.50 & 0.45 & 0.45 & 0.44 & 0.41 & 0.40 \\
N=10                & 0.68 & 0.49 & 0.44 & 0.43 & 0.42 & 0.39 & 0.37 \\
N=11                & 0.65 & 0.44 & 0.43 & 0.40 & 0.40 & 0.38 & 0.37 \\
\hline $\delta t$   & 0.002 & 0.005 & 0.007 & 0.008 & 0.009 & 0.010 & 0.015 \\ \hline
\end{tabular}
\caption{\em
Computational Time (CT) for the minimum residual - projective CG algorithm.
The table is normalized to 1 at $\delta t=0.010$
for N = 1 ( $\chi_{trial}=\chi_{last}$). 
\label{tab:tavola6}}
\vskip 1cm
\end{table}

Our simulations show that the Minimal Residual Extrapolation algorithm
followed by the PCG algorithm gives us an extra $30\%$
improvement over the Minimal Residual Extrapolation followed by the
standard CG. In general the actual improvement of the performance
depends crucially on the dot product versus Matrix-Vector product
speed that a given computer can achieve.

\pagebreak

\vskip 0.7cm

{\bf VIII. Conclusions}

\begin{figure}
$$
\epsfxsize=15.5cm
\epsfysize=14.5cm
\epsfbox{ 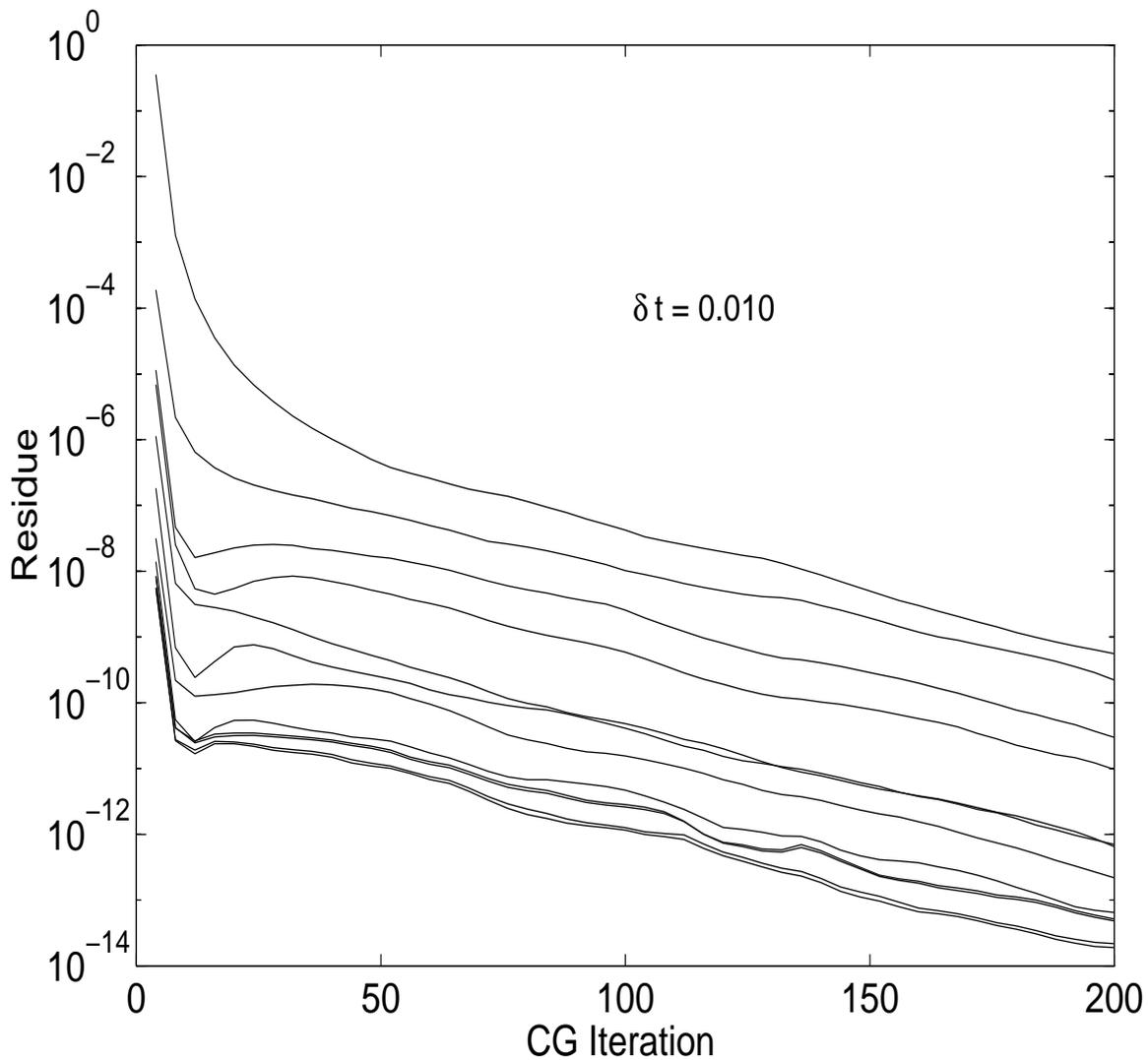} 
$$
\caption{ Example of convergence of the CG as a function of the
number of extrapolation vectors for a single lattice and
$\delta t=0.010$. The method used was Minimal Residual Extrapolation
 and the number of
vectors varies from $0$ (top line) to $11$ (bottom line). \label{fig:CGconv}}
\end{figure}

Let us summarize our perspective on the problem of accelerating a time
sequence of conjugate gradient inverters. The chronological method has
offered significant performance improvement, but at the same time the
results are tantalizing.  Just by starting from the old solution, the
residue is reduced by 4 orders of magnitude to order $10^{-4}$,
relative to the residue with $\chi=\varphi$. Then if we look at
Figure~\ref{fig:CGconv}, we see that the residue is reduce further by
5 orders of magnitude using 10 additional past solution vectors in our
Minimal Residual Extrapolation method. Finally, the first 10 CG
iterations accounts for another 2 orders of magnitude. However
accurate reversibility ultimately requires us to reduce the residue by
another 4 more orders. This last 4 orders of magnitude takes several
hundred additional vectors in the CG iterations. We are both intrigued
and frustrated by the observation that we can reduce the residue by 11
orders of magnitude in a 20-30 dimensional vector space, but then the
standard CG iterations requires hundreds of additional search
directions to accomplish the remaining 4 orders of magnitude needed to
satisfy adequately the reversibility constraint. It is tempting to
hope that further improvements can be made on this last 4 orders of
magnitude.

We have emphasized the analytic properties of $\chi(t)$ because we
believe it may suggest ways to understand and further improve the
chronological method.  For example, the failure at fixed $\delta t$ to
improve the polynomial extrapolation by increasing indefinitely the
number of terms is probably a signal of nearby singularities.  Our
success so far is probably due to the slow evolution of the low
eigenvalues of the Dirac operator. Thus we are in essence taking
advantage of ``critical slowing down'' in the HMC algorithm to
accelerate the Dirac inverter.  However, there may well be other
vectors (besides the final solution) in the nearby past iteration that
can better exploit the slow evolution of our matrix. For example, the
last CG routine, which is closest in time to our present inverter,
itself generates many A conjugate search vectors, that may be more
useful than the older solutions exploited in our MRE method. In
addition this subspace of past vectors might be used not just as a way
to arrive at an initial guess, but also as a way to accelerate (or
``precondition'') the iterative process itself.  In this spirit the
Projective CG method presented in section VII was included to
illustrate how in principle a set of past vectors can be exploited to
accelerate the convergence of the CG algorithm itself.  We are
currently studying such chronological preconditioners more
systematically.

In conclusion, we found that the procedures presented here, and in
particular the Minimal Residual Extrapolation method reduces by a
factor of about 2-3 the mean number of conjugate gradient iterations
required to move in phase space for a fix molecular dynamics time.
This is achieved with negligible extra computational time, at the
expense of memory.  Consequently when sufficient memory is available
for storing the past solutions the Dirac inverter, our Chronological
Inversion method certainly provides one more useful trick for more
efficiently generating full QCD configurations.

{\bf Acknowledgments:} We would like to thank D. Castanon,
C. Evangelinos and C. Rebbi for many helpful discussions. This work
was supported in part by funds provided by the U. S. Department of
Energy under grand DE--FG02--91ER400688, Task A, and under cooperative
agreement \#DF-FC02-94ER40818.

\pagebreak

%% FIGURES %%

%%%%%%%%%%% Reversibility %%%%%%%%%%%%%

%%%%%%%%%%%%%% Minimal Residual %%%%%%%%%%%%%%%%%%%%%%

%%%%%%%%%%%%%%% Polynomial %%%%%%%%%%%%%%%%%%%%%%

%%%%%%%%%%%%%%%%%%%% Projective %%%%%%%%

%%%%%%%%%%%%%% CG convergence %%%%%%%%%%%%%%%%%%%%%%%%%%%

\end{document}